\documentclass[a4paper,twoside]{article}
\newcommand{\affil}[1]{$^{\rm #1}$}
\usepackage[authoryear]{natbib}
\bibpunct{(}{)}{;}{a}{}{,}
\usepackage{graphicx}
\date{} 
\newcommand{\kms}{\mbox{km\,s$^{-1}$}}
\newcommand{\kmsM}{\mbox{km\,s$^{-1}$\,Mpc$^{-1}$}}

\newcommand{\la}{\mathrel{\mathchoice {\vcenter{\offinterlineskip\halign{\hfil
$\displaystyle##$\hfil\cr<\cr\sim\cr}}}
{\vcenter{\offinterlineskip\halign{\hfil$\textstyle##$\hfil\cr
<\cr\sim\cr}}}
{\vcenter{\offinterlineskip\halign{\hfil$\scriptstyle##$\hfil\cr
<\cr\sim\cr}}}
{\vcenter{\offinterlineskip\halign{\hfil$\scriptscriptstyle##$\hfil\cr
<\cr\sim\cr}}}}}
\def\farcs{\hbox{$.\!\!^{\prime\prime}$}}
\def\fo{\hbox{$.\!\!^{\rm \circ}$}}
\title{\large\bf\flushleft Distance Measurements 
and Stellar Population Properties via \\ Surface Brightness Fluctuations}
\author{\parbox{\textwidth}{\flushleft
\vspace{-0.5cm}
{\it Alexander Fritz\affil{A,B,C}}\\
\vspace{0.4cm}
{\small \affil{A}\,Gemini Observatory, 670 N.\ A'ohoku Place, Hilo, HI 96720, USA}\\
{\small \affil{B}\,Istituto Nazionale di Astrofisica -- Istituto di Astrofisica
Spaziale e Fisica Cosmica Milano, Via E. Bassini 15, 20133 Milano, Italy}\\
{\small \affil{C}\,VIPERS Fellow; Email: afritz@iasf-milano.inaf.it}}}
\begin{document}
\twocolumn[
\begin{changemargin}{.8cm}{.5cm}
\begin{minipage}{.9\textwidth}
\vspace{-1cm}
\maketitle
%
%
\small{\bf Abstract:}
Surface Brightness Fluctuations (SBFs) are one of the most powerful techniques
to measure the distance and to constrain the unresolved stellar content of
extragalactic systems. 
For a given bandpass, the absolute SBF magnitude $\overline{M}$ depends on
the properties of the underlying stellar population. Multi-band SBFs allow
scientists to probe different stages of the stellar evolution: 
UV and blue wavelength band SBFs are sensitive to the evolution of stars
within the hot Horizontal Branch (HB) and post-Asymptotic Giant Branch
(post-AGB) phase, whereas optical SBF magnitudes explore the stars within the
Red Giant Branch (RGB) and HB regime. Near- and Far-infrared SBF luminosities
probe the important stellar evolution stage within the AGB and
Thermally-Pulsating Asymptotic Giant Branch (TP-AGB) phase.
Since the first successful application by Tonry and Schneider,
a multiplicity of works have used this method to expand the
distance scale up to 150 Mpc and beyond.
This article gives a historical background of distance measurements, 
reviews the basic concepts of the SBF technique, presents a broad sample
of these investigations and discusses possible selection effects, biases, and
limitations of the method. In particular, exciting new developments and
improvements in the field of stellar population synthesis are discussed that
are essential to understand the physics and properties of the populations in
unresolved stellar systems. Further, promising future directions of the SBF
technique are presented. With new upcoming space-based satellites such as 
{\em Gaia}, the SBF method will remain as one of the most important tools 
to derive distances to galaxies with unprecedented accuracy 
and to give detailed insights into the stellar content of globular clusters
and galaxies.


\medskip
\date{\textit{{\small CSIRO Publishing - PASA; Received 2011 December 22, 
accepted 2012 March 26}}}
\medskip
\end{minipage}
\end{changemargin}
]
\small

\section{Introduction}\label{one}

\subsection{Early Concepts}

One of the oldest and most challenging problems in astronomy is the
measurement of distances to any astrophysical object on the sky. 
Although this task appears to be very basic in theory,
it represents a major challenge in practice.

Aristarchus of Samos (310-230 BC) is probably the first astronomer to measure
the distance between the sun and moon using trigonometry. In 265 BC he estimated
the angle between moon-earth-sun at the point where the moon is exactly
half full to $87^{\circ}$. Using the relation
$r/R=\cos 87^{\circ}$, with $r$ and $R$ being the distance to the moon and sun,
respectively, the distance to the sun should be 19 times the distance between
earth and moon. Today we know that the exact angle is only sightly different
($89\fo85$). However, not surprisingly the effect on the derived distance is
huge, with the distance of the sun from the earth being 390 times the distance
between earth and moon. Several other Greek astronomers, like Eratosthenes,
Hipparcos, and Ptolemy, not only continued to measure the sizes of the earth and
moon but also estimated the relative distances to the planets as far out as
Saturn.

\subsection{Important Steps Towards the \newline
Cosmological Distance Ladder}\label{cont}

Significant advances regarding the astronomical distance scale had to await
the 16th century, where Nicolaus Copernicus (1473-1543) introduced for
the first time almost correct relative distances of the planets from the sun.
In 1672, John Flamsteed (1646-1719)
and independently Giovanni Domenico Cassini (1625-1712) measured the distance
between the earth and sun to $\sim$140 million km and underestimated this
distance only by 7\%. 
At the same time, Sir Christopher Wren (1632-1723) proposed
that nebulae might be distant star systems with the light of their individual
stars being blurred into a milky, continuous glow (the `island universes' theory
of nebulae). However, it took until the 20th century to take the next important
step concerning the distance scale in establishing the distance to these 
(extragalactic) spiral nebulae through the discovery and application of
variable Cepheid stars in the Magellanic Clouds (MCs) by Henrietta Swan Leavitt
(1868-1921). Harlow Shapley (1885-1972; Shapley 1919) investigated Milky Way
Globular Clusters (GCs) and measured their distance with pulsating Cepheid stars
that were calibrated using the Period-Luminosity ($P-L$) relationship for galactic
Cepheids as established by Leavitt (1908).
Assuming that the GCs are part of our own galaxy with an increased concentration
towards the direction of Sagittarius, Shapley concluded that the central core of
the Milky Way must be towards Sagittarius too. This correct picture, though
having a wrong diameter for the Milky Way (100 kpc) due to underestimation
of the contribution of interstellar extinction (dust), yielded in 1920 to the
famous {\em great debate} between Heber Doust Curtis (1872-1942) and Harlow
Shapley. Heber Curtis, himself a strong supporter of the `island universe'
theory, resolved individual stars in spiral nebulae
that have created novae and concluded from a comparison with novae in our own
galaxy that the novae in spiral nebulae need to be located far away.
The final evidence that spiral nebulae are
extragalactic systems was given later by Edwin Powell Hubble (1889-1953;
Hubble 1926). Using new Cepheid discoveries, that were calibrated to Leavitt's
$P-L$ relationship of galactic Cepheids, he showed that M31 and M33 were
far more distant than the Milky Way. Additional support was provided by Robert
Julius Tr\"umpler (1886-1956), who discovered the extinction to open (galactic)
star clusters, and Vesto Malvin Slipher (1875-1969), who from 1912 onwards 
measured radial velocities of spiral galaxies with large recession velocities
from the Doppler speed of spectral lines. In 1929, Hubble found a linear
correlation between the apparent distances of galaxies $d$ and their recession
velocities $v_{\rm pec}$ of form $v_{\rm pec}=H_0\cdot d$
(also called the Hubble law, Hubble 1929, see Section~\ref{hst}).
The convincing conclusion that there indeed exists a relation between redshift
and distance was given two years later using new observational data
(Hubble \& Humason 1931). However, this major advance in
understanding did not suggest an expansion of the universe per se.
The hypothesis of an expanding, homogeneous and isotropic universe that is
still valid today was introduced with the mathematical works of Aleksandr
Friedmann (1888-1925),  Georges Lema{\^i}tre (1894-1966, Lema{\^i}tre 1927),
and Howard Percy Robertson (1903-1961) who all contributed independently to
the classical solution of the Einstein field equations.
Using the Hubble law and the assumption that the
expansion of the universe was slower at earlier times, the age of the universe
today should be 1.8 Gyr and the Hubble constant $H_{0}=450$~\kmsM.

With the application of new or improved techniques in measuring
distances to extragalactic objects, the extragalactic distance scale was
subsequently extended to larger distances. Walter Baade (1893-1960; Baade 1944)
was the first to separate the stellar populations of spiral galaxies and GCs and
derived a much smaller $H_{0}$ of 250~\kmsM. Allan Rex Sandage (1926-2010)
discovered that the `bright stars' seen by Hubble are actually H{\small II}
regions that consist of star groups and clouds of ionized hydrogen. 
Sandage's $H_{0}$ of $\sim$75 km\,s$^{-1}$ Mpc$^{-1}$
corresponds to an age of the universe of $\sim$13 Gyr. 
For a recent progress report on measuring
the Hubble constant to a precision of 5\% using the most accurate distance
indicators, the reader is referred to Freedman \& Madore (2010),
whereas observational pro\-bes to constrain the cosmic acceleration are presented
in Riess et al. (2011) and Weinberg et al. (2012).

\begin{figure*}[t]
\begin{center}
\includegraphics[angle=0,width=6.in]{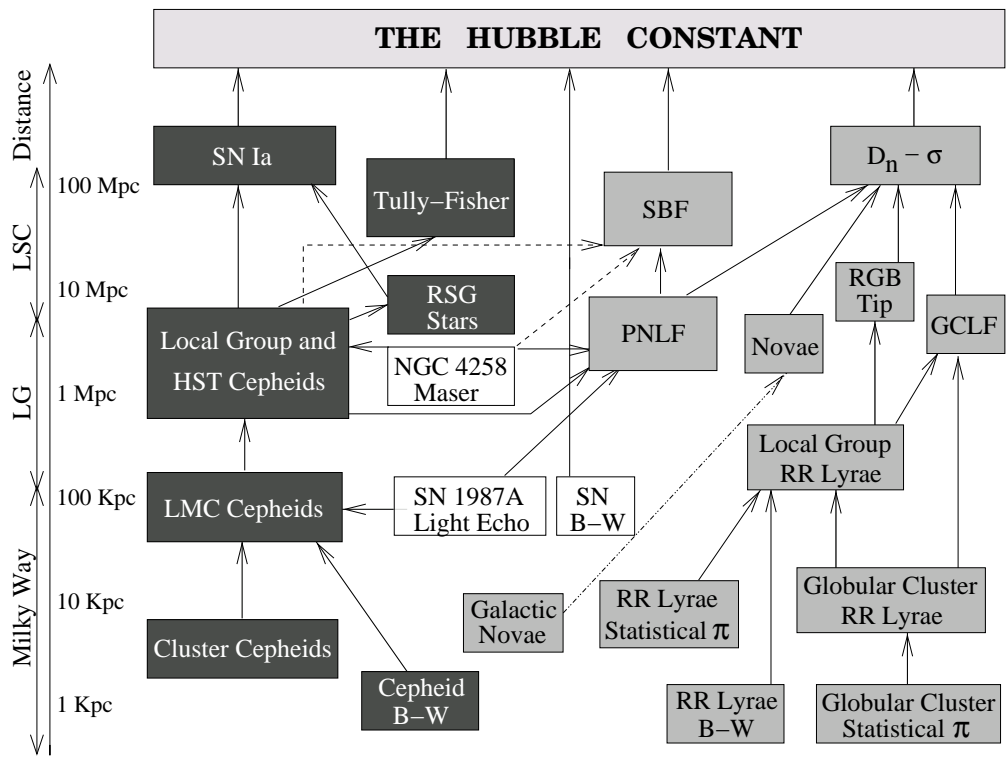}
\vspace{0.2cm}
\caption[SBF gradients]{\small{
The extragalactic distance ladder. The dark boxes represent techniques useful
in star-forming stellar systems (young Population I stars), grey-filled boxes
denote methods that can be applied to quiescent stellar systems (evolved
Population II stars), whereas open boxes give distance determinations using
geometric techniques. Less-certain calibration steps are indicated as dashed
lines. The distance increases logarithmically from bottom to top.
Due to space constraints, the following abbreviations have been used: 
Baade-Wesselink method (B-W), Globular Cluster Luminosity Function (GCLF),
Local Group (LG), Local Super Cluster (LSG),
Planetary Nebula Luminosity Function (PNLF), Red Giant Branch (RGB),
Red Super Giant (RSG), Supernova (SN), Surface Brightness Fluctuations (SBF), 
Parallax ($\pi$). The figure was adapted from Ciardullo (2006).}}
  \label{dladder}
\end{center}
\end{figure*}

Figure~\ref{dladder} displays the extragalactic distance ladder with the most
important standard candles in astronomy. The distance increases
logarithmically from bottom to top, starting with our Milky Way, across
the Local Group (LG) and extending towards the Local Super Cluster (LSC).
Each technique represents an individual step in the distance ladder that is
calibrated with a previous method of less-distant objects. The ultimate goal
is a precise measurement of the Hubble constant that also gives information
about the adopted galactic luminosity function and allows us to test models for 
dark energy (see also Section~\ref{con}).

%
%

\subsection{A Brief History of SBFs}\label{his}

Early attempts in the field of Surface Brightness Fluctuations (SBFs)
were conducted by Baade (1944), who resolved Population II stars in
galaxies of the Local Group. Individual star fluxes in external galaxies
constitute a significant part of the stellar luminosity function of red giant
stars and yield, besides the distance of the galaxy, substantial information
on the stellar populations. Soon thereafter, Baum \& Schwarzschild (1955)
obtained observations of M31 and M32 (NGC 205) close to the limit of the
resolution (historically called {\em `incipient resolution'})
and investigated their stellar populations using the ratio
of the number of resolved stars to the total integrated light (the so-called
`count-brightness ratio'). The empirical ratio of $N/L_{V,{\rm tot}}$
is a useful proxy because it is sensitive to differences in the average
stellar populations. The brightness was expressed in terms of visual ($V$-band)
luminosities, but is in principle applicable to all wavelengths. Other works
in similar directions followed. Mould et al. (1983; 1984) found similar
luminosity functions for the dwarf ellipticals NGC 147 and NGC 205, whereas
Pritchet \& van den Bergh (1988) and Freedman (1989) measured the tip of the red
giant branch via colour-magnitude diagrams for M31 and M32, respectively. 
However, individual stars can only be resolved for galaxies within the Local Group,
thereby limiting the power of distance measurements considerably. Unfortunately,
this restriction will not significantly change with the introduction of the
next generation of large telescopes, see further Section~\ref{fut}.
Nevertheless, this problem of measuring individual star fluxes without resolving
them can be bypassed if a characteristic flux of a stellar population is
determined. For example, for the red populations of globular clusters this
average characteristic flux can be best measured at the bluest wavelengths
where the red fluctuations are smallest.
An alternative way to derive fluctuation amplitudes is by using
a histogram of pixel amplitudes (Baum 1986, 1990). This statistical analysis of 
pixel brightness variations is a variant of the `count-brightness ratio' method
and can be used to establish the upper end of the Hertzsprung-Russell diagram
and through a match between simulations and real data can
estimate distances as far as the Virgo galaxy cluster.

In the late eighties, Tonry and Schneider invented a new way to quantify
and measure surface brightness fluctuations to determine extragalactic
distances (Tonry \& Schneider 1988). The basic technique of SBFs consists of
spatial pixel-to-pixel brightness variations, so-called fluctuations, which
arise due to the varying number of stars within the pixels in a high
signal-to-noise-ratio ($S/N$) Charge-Coupled Device \newline (CCD) image of a stellar
system, like a (globular) star cluster or an early-type galaxy or spiral bulge.
The details of the theory and the basics of the method are described in
Section~\ref{met}.

Over the past decades, several articles have discussed some aspects of the SBF
technique (e.g., Tonry et al. 1990; Tonry 1991; Jacoby et al. 1992;
Tonry et al. 1997). However, there has been only one review on the subject,
which concentrated mainly on the $I$-band (Blakeslee et al. 1999).  

This article presents an updated critical review of progress in the application
of the SBF method. In particular, the present work focuses on new developments
in this field, like the use of near-infrared (NIR) photometric bands or recent
theoretical model improvements. The article is organised as follows.
In Section~\ref{met}, the origin and technique of surface brightness
fluctuations is explained in detail. Section~\ref{obs} gives a short overview of
observational attempts in determining SBF distances from both the optical and
NIR perspectives. Selection effects and biases that may alter the
accuracy of the SBF method and the application of SBF as a distance indicator
are presented in Section~\ref{dis}. In Section~\ref{mod}, SBFs are
discussed from a theoretical point of view. Further,
this section describes the recent progress obtained in stellar
population synthesis modelling. An alternative way of calibrating SBF
observations is presented in Section~\ref{ncal}. Concluding remarks are drawn
in Section~\ref{con}. Finally, Section~\ref{fut} outlines some forecasts for
the future.

\section{The Method}\label{met}

The basic idea behind the SBF 
technique is quite strai\-ghtforward: the fluctuations in unresolved star
clusters of the Milky Way and other galaxies appear, due to the
Poisson statistics (which is the counting statistics of discrete stars),
to be clumpy and mottled, whereas in more distant objects the fluctuations
appear more smooth. These spatial brightness variations, which are 
distance-dependent in amplitude and varying for a given resolution element,
are proportional to $\bar{f}\sqrt{N}$, where $\bar{f}$ denotes the mean flux
per star and $N$ is the mean number of stars per pixel.
Note that these values are summed over all stars of a stellar system.
The mean intensity per resolution element is $N\bar{f}$; therefore the
difference between the spatial variance and the observed mean results in the
average flux per star $\bar{f}$, which decreases inversely with the square of
the distance $d^{-2}$. $\bar{f}$ is the average flux of the underlying stellar
population, weighted to its luminosity, and corresponds for evolved stellar
populations roughly to the flux of a typical old giant star
between spectral type K to M within the Hertzsprung-Russell diagram.
If the mean absolute magnitude $\overline{M}$
is known, it is possible to determine the distance of the target object.
Inversely, knowing the distance, $\overline{M}$ can be established, which 
itself yields information on the stellar population of a galaxy.

The average stellar luminosity $\overline{L}$ of a stellar system (star cluster
or galaxy), which is summed over all stars (see equation~\ref{LF}),
is defined as $\overline{L}=4\pi\bar{f}\,d^2$. $\overline{L}$ is
weighted towards the brightest stars of a specific population.
For the typical old, evolved and metal-rich stellar
populations of elliptical galaxies, these are cool, luminous, Red
Giant Branch (RGB) stars (3.2$\la$ log($L_{\rm bol}/ L_{\odot}$)$\la$ 4.2)
and Thermally-Pulsating Asymptotic Giant Branch (TP-AGB) stars.
Since (TP-)RGB stars have a red colour, the SBF magnitudes are red;
ellipticals typically display $\overline{m}_I-\overline{m}_K\approx4.20\pm0.10$
(Jensen et al. 1998). The preferred photometric bands for SBF observations are
either the red ($VRI$) or the NIR filter bandpasses ($JHK$). NIR filters are
preferred over optical wavelength bands, owing to two advantages in
particular: {\it (i)} the SBF signal is dominated by red, luminous giant stars,
and  {\it (ii)} there are smaller extinction corrections due to lower dust
absorption in the stellar systems. However, the Kron-Cousins $I$-band supersedes
the NIR-bands because of its relative insensitivity to spatial variations in the
stellar populations.

%
%
\begin{figure}[!t]
\begin{center}
\includegraphics[angle=0,width=3.0in]{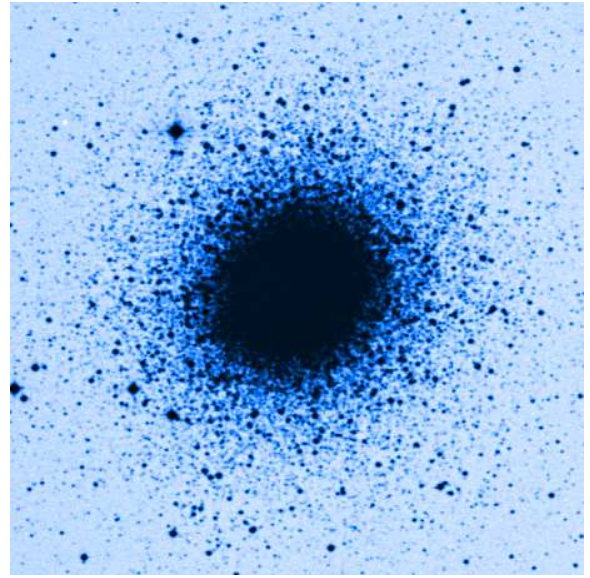}
\vspace{0.2cm}
\caption[SBF mottling]{\small{
A high $S/N$ exposure of the galactic globular cluster M2. The composite
ESO/Digitized Sky Survey 2 (DSS2)\footnotemark[1] image clearly shows an 
obvious strong lumpiness and mottling that is caused by the dominant evolved
giant star population. The same mechanism works for other stellar systems,
such as galaxies, but is less evident because of their more complex stellar
content and greater distance.}}
  \label{GC}
\end{center}
\end{figure}
\footnotetext[1]{ESO Online DSS2:~http://archive.eso.org/dss/dss}

An observational example for the effect of spatial brightness variations
is given in Figure~\ref{GC}. A high $S/N$ image of the galactic globular cluster
M2 shows very nicely a strong lumpiness and mottling caused by the dominant
evolved giant star population. In principle, this mechanism also works for
galaxies, but for enhancing effects a GC was chosen. In a first step, a
CCD detector measures the total flux per pixel. From this total flux
the average flux per pixel (which is also known as surface brightness)
and the root-mean-square (rms) variation of the flux from pixel-to-pixel
can be derived. However, it is impossible to distinguish the two stellar
systems by their average flux per pixel, because the number of stars
per pixel of a resolution element increases with the distance $d^{2}$,
whereas simultaneously the flux per star decreases inversely with the 
square of the distance ($1/d^{2}$). 
Stars cannot be resolved individually; only a characteristic (mean) flux
per pixel can be established. If the number of detected photons within a
resolution element is larger than the projected number of stars within this
area, the fluctuations are proportional to the square root of the number of
stars; i.e., the variations follow $\bar{f}\sqrt{N}$, with $\bar{f}$ being the
average flux per star and $N$ the average number of stars per pixel. Thus, the
variance of fluctuations $\sigma^{2}_{f}$ is derived from the square of the
fluctuations from pixel-to-pixel as $\sigma^{2}_{f}= \bar{f}^{2}\,N$, where
$N\bar{f}$ denotes the mean flux per pixel.
The average flux per star is determined from the ratio between the
fluctuation variance and the average flux per pixel as
\begin{eqnarray}
\bar{f}\,=\,\frac{\sigma^{2}_{f}}{\overline{f}_{pix}}\,=\,
\frac{\overline{f}^{2}\,N}{\overline{f}_{pix}}. 
\label{fucflux}
\end{eqnarray}
$\bar{f}$ is the average flux of the underlying stellar population, weighted to
its luminosity, and corresponds for evolved stellar populations 
roughly to the flux of old RGB stars.

%
%
\begin{figure*}
\begin{center}
\includegraphics[angle=0,width=6.in]{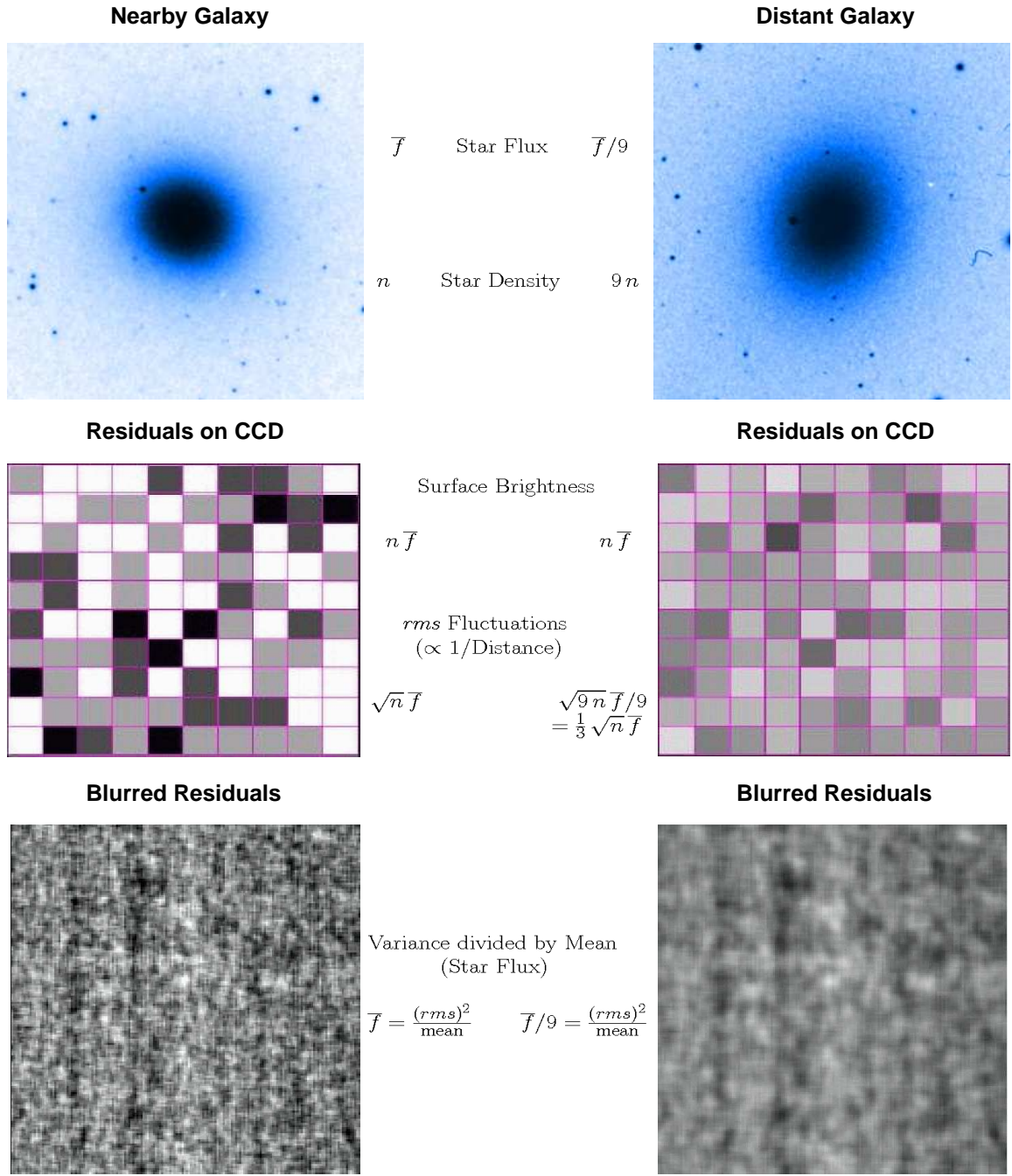}
\caption[SBF method]{\small{Sketch of the Surface Brightness Fluctuation
  method. A \textit{nearby} galaxy (G1, left\footnotemark[1]) is compared to a
  \textit{distant} galaxy (G2, right\footnotemark[1]) with a three times larger distance than the
  nearby stellar system. Note the slightly mottled structure seen in the
  outer parts of galaxy G1. The middle panels represent the galaxy-subtracted
  SBF residual view as seen on a CCD chip. One square in the image corresponds
  to a single CCD pixel. The bottom panels show the galaxy-subtracted SBF
  residuals convolved with the observed seeing to simulate the blurring caused
  by the earth's atmosphere. See text for a detailed description.}}
  \label{sbfketch}
\end{center}
\end{figure*}

A sketch of the SBF technique is illustrated in Figure~\ref{sbfketch}.
Let us compare two stellar systems, a \textit{nearby} galaxy
(G1, left panels of Figure~\ref{sbfketch}) and
a second more \textit{distant} galaxy (G2, right panels of Figure~\ref{sbfketch}),
three times as distant as the first one. Let us now assume that
for the nearby G1 the average number density is 100 stars/pixel.
The fluctuations with rms variations from pixel-to-pixel (rms fluctuations)
are therefore 10\% of the mean signal. At different regions in the galaxy, the
fluctuations vary as the square root of the underlying local mean galaxy
brightness. Therefore, there is a proportionality constant between the rms
fluctuations and the square root of the mean surface brightness, which is
directly related to the number of stars present: $rms\propto \sqrt{N}$$\bar{f}
\propto 1/d$. If we now consider the distant galaxy, G2 has an average number
density of 900 stars/pixel. At the same time, the star flux is decreased by a
factor of 9 (the surface brightness remains constant), thus the galaxy contains
only one-third of the rms fluctuations from pixel-to-pixel (3.3\% of the mean 
signal). In the second galaxy, G2, different regions in the galaxy will follow
a linear relationship between the rms fluctuations and the square root of the
galaxy flux. A comparison between G1 and G2 yields the surface brightness as a
constant. As the proportional constant of a stellar system is inversely
proportional to the distance ($1/d$), the constant of galaxy G2 is one-third of
that for the nearby galaxy G1.

From a theoretical point of view, the SBF method relies on using the ratio of
the second moment to the first moment of the stellar luminosity function (LF)
of the galaxy as:
\begin{equation}
\overline{L}\equiv\frac{\sum_i n_{i}\,L_{i}^{2}}{\sum_i n_{i}\,L_{i}},
\label{LF} 
\end{equation}
with $n_i$ being the number of stars of spectral type $i$ and
luminosity $L_i$.
The mean fluctuation luminosity $\overline{L}$ depends on the stellar population
and the galaxy colour itself depends on the underlying population.
It is important to mention that the relation in equation~\ref{LF} is very
insensitive to the uncertain faint end of the LF. Assuming a simple power-law
luminosity function, the fluctuation luminosity $\overline{L}$ 
scales linearly with the maximum luminosity of the stars $L_i$.

The primary utility of SBFs as an extragalactic distance
indicator will be evaluated in Section~\ref{dis}.
Since the data collected for SBF analysis can also be used to determine the
surface brightness, which provides information about the first moment of the
stellar LF, the surface brightness therefore allows a measure of the stellar
content {\em solely} from the integrated flux. The utility of SBFs as stellar
population tools will be discussed in Section~\ref{grad}.
Further, as the SBFs also depend on the second moment of the stellar LF, the
SBF signal is more sensitive to the most luminous stars in a stellar system. In
case of elliptical galaxies, these stars are evolved cool giant stars.
The application of SBFs as a constraint on the evolution of evolved stellar
populations will be presented in Section~\ref{mod}.

\subsection{Types of Fluctuations}\label{fluc}

Within a single CCD exposure of a stellar system there are a number of
pixel-to-pixel fluctuations and the individual sources can basically be
divided into three groups: (1) intrinsic fluctuations from the target of
interest itself (e.g., a galaxy), (2) fluctuations from other objects, and (3)
fluctuations caused from the instrumentation.

\textit{Intrinsic fluctuations} are sources that account for additional
contributions to the basic galaxy fluctuation signal and are part of the
galaxy or its environment (e.g., GCs, H{\small II} regions, planetary nebulae,
satellite systems (dwarf galaxies), etc). Moreover, the SBFs
we are interested in and want to measure are included in this group
too. Other fluctuations arise from \textit{unwanted sources} (primarily
foreground stars or background galaxies) that are on the same exposure as the
target of interest. The third group of fluctuations originates from the
instrumental setup used (telescope and optics)
and the detector itself (readout noise of the CCD camera, photon
shot-noise) and from cosmetic artefacts (cosmic rays, traps, defect columns, 
noise from the counting statistics of the flatfield exposures, or
residual flattening problems after the flatfield correction). 
Additional Poisson noise is generated 
from the number of detected photons in each pixel of the target itself.

Patchy (dust) obscuration results in images that are mottled or partly obscured
and affects the flux of a stellar system by creating additional variance. Smooth
(dust) obscuration can reduce the derived flux and the variance we seek to
measure. By restricting the SBF analysis to elliptical and S0 galaxies, the
problems of patchy dust obscuration are largely reduced. Moreover, these
early-type galaxies exhibit such high velocity dispersions that only dense
(c)lumps endure in the fluctuation flux measurements, hence limiting
possible pixel-to-pixel correlations from gravitational clumping. However,
using high-$S/N$ images with sufficient resolution at blue wavelengths ($U$ or
$B$-band) the contamination of dust can be firmly established and hence the
SBF technique can be extended to early-type spiral galaxies or bulge-dominated
galaxies.

There are two main factors of the fluctuations cau\-sed by the 
instrumentation used: The readout noise and the device photon shot-noise. The
variance of the readout noise is given by a relation between the inverse CCD
gain $a$ (in units of electrons per analog-to-digital unit (ADU)) and the CCD
readout noise $N_{R}$ (in units of electrons) as (Tonry \& Schneider 1988):
\begin{equation}
\sigma_{R}^{2}=\frac{N_{R}^{2}}{a^{2}} \label{readout}.
\end{equation}
The formula of the variance of the readout noise is composed of the
contributions from the inverse CCD gain $a$ and the average total signal
$\overline{c}(x,y)$ at the point $(x,y)$ on the CCD chip in ADU. For the mean
total signal, the bias level was subtracted and the detector response and
pixel-to-pixel variations (flatfield) were corrected, whereas the mean sky
brightness has not been removed. The variance of the photon shot-noise,
which is the noise due to the photon counting statistics, is defined as
\begin{equation}
\sigma_{P}^{2}=\frac{\overline{c}}{a} \label{photon}.
\end{equation}
The average total signal $\overline{c}$ is denoted by the relation 
$\overline{c}=\overline{g}+s$, with $\overline{g}(x,y)$ being the mean signal
of the stellar system at the point $(x,y)$ in ADU and $s$ being the sky flux.

All these sources of instrumental related noise are described by a white power
spectrum and can therefore be separated from the intrinsic fluctuations from
the stellar component of the target system, which are represented by a power
spectrum of a point-spread-function (PSF); see further Section~\ref{pract}.
Intrinsic fluctuations produced from sources other than the target object 
(e.g., stars, globular clusters, faint background galaxies) are more difficult,
because these sources are characterized by a similar power spectrum as the
spatial luminosity fluctuations that we are interested in.
Further, the variance signal of the luminosity fluctuations could be
corrupted by the presence of spiral arms or star forming regions, which 
would invalidate the general assumption that adjacent pixels are independent
samples of the average stellar population.

\subsection{SBF Signal Measurement}\label{pract}

In the following, the basic procedures involved in measuring a fluctuation flux
are presented. The details of the complete data reduction steps and analysis
tools are beyond the scope of this review. The interested reader is therefore
referred to Tonry et al. (1990, hereafter TAL90); Sodemann \& Thomsen (1995);
Blakeslee et al. (1999); Fritz (2000, 2002). 

Distance determinations using SBFs are based on two individual steps
that are linked together:
{\it (i)} measurement of a fluctuation flux, and {\it (ii)} conversion to an
absolute distance by assuming a calibrated, absolute fluctuation luminosity.
There are several important facts regarding the SBF method:
\begin{itemize}
\item Preferred targets are dust-free systems, like elliptical (E)
and S0 galaxies, spiral bulges or globular clusters.
\item The total integration time must be sufficiently long enough to collect 
$\ge$5 photons per source with apparent magnitude $\overline{m}$. Moreover,
the PSF must be stable and uniform across the field during the collection
of the observations. In particular, the telescope exposure time must exceed
the point at which the photon shot noise per pixel is smaller than
the intrinsic SBFs. This breaking point is given by the time at which
1~photo\-electron is collected per giant star of brightness
$\overline{m}$, with the integration time defined as 
\begin{equation}
T\,{\rm [sec]}\,=\,(S/N)\,10^{0.4(\overline{m} - m_1)},  \label{expt}
\end{equation}
where $S/N$ represents the required signal-to-noise ratio and $m_1$ denotes
the magnitude for 1~detected photo\-electron per second on the CCD detector.
Typically, $\sim$5-10~$e^-$ for a star with $\overline{m}$ are sufficient to
reach a $S/N \sim 5-10$, and an observation strategy with several dithered
exposures is recommended to limit the impact of possible image gradients,
fringe patterns, or CCD defects. The uncertainties in $\overline{m}$ are
defined as $\delta \overline{m}=0.03\,
(\frac{{\rm FWHM}}{{\rm arcsec}}\,\times\,\frac{d}{1000\,km~s^{-1}})^2$,
with the first error term being the full-width-at-half-maximum (FWHM)
of the PSF and the second the distance $d$.
\item The photometric $I$-band is a useful tool for distance measurements
because the absolute fluctuation magnitude $\overline{M}$ is even
brighter than the high sky background level in the NIR wavelength
regime. The usage of this band also limits the impact of dust absorption.
Moreover, for cluster galaxies there exists an empiric relationship of
$\overline{M}_I$ with the mean galaxy colour with a small scatter 
($\sim$0.07 mag), which can be used to derive distances (see further
Section~\ref{cal} and equation~\ref{T01}).
\item The calibration of the zeropoint of the absolute fluctuation magnitude
$\overline{M}_{I}$ can be based on theoretical stellar population synthesis
models (SBF acts as a primary distance indicator) or derived from an empirical
relationship (SBF is used as a secondary distance indicator). The empirical
calibration is based on observations of galactic GCs or
LG galaxies. The methods using predictions from stellar population
prescriptions and galactic GCs are expected to be consistent at least to
the 10\% level.
\end{itemize}

In practice, the basic procedure to measure a fluctuation flux is to perform 
high $S/N$ observations and data reductions in such a way as to obtain a smooth
and uniform image that is based on a precise photometric calibration. To ensure
a reliable calibration, a significant fraction of the observing time must be
devoted to photometric standard stars. 
As can be seen from equation~\ref{expt}, the fluctuation signal increases
proportional to the exposure time and the ideal assumption is to obtain 
$\ge$10~photo\-electrons for a star of $\overline{m}$ in order to leave the
regime of photon-counting (shot-noise) statistics and enter the regime
of star-counting statistics. Therefore, the integration time depends only on the
SBF magnitude $\overline{m}$ and the detector sensitivity, but is independent
of the size of the target object. Hence, in the absence of a sky background 
all points in a stellar system $p_i$ are described by the same ratio of
variance of fluctuations to variance of photon-counting statistics as
$p(x,y)\propto\sigma^{2}_{f}/\sigma_{P}^{2}+(1+s/\overline{g})$.

Once the initial data reduction (including cosmic ray removal) and photometric
calibration is complete, obvious point sources, background galaxies, any regions
contaminated by CCD defects or dust are masked out and the sky background level
in the outer districts of the stellar system is estimated using a $r^{1/n}$
law profile ($n=4$ for E+S0 galaxies) plus constant offset (which allows both
measurements of the mean galaxy colour and $\overline{m}$ as function of
radius). In the next step, elliptical isophotes are fitted to the masked
sky-subtracted galaxy image and a smooth galaxy model is constructed. This
galaxy model is then subtracted from the masked image and corrected for
large-scale background deviations (which affects only the low wave numbers in
the image power spectrum, which are excluded from the determination of
$\overline{m}$) to yield a residual image with various fluctuation
contributions. From this residual image the Fourier power spectrum is
constructed and the mean variance measured.

Before the determination of the average fluctuation variance, the
contributions of fluctuations must be accounted for.
Foreground stars, background galaxies, and GCs are identified and classified 
with an automatic photometric program to a specific completeness level
(e.g., Sodemann \& Thomsen 1995; Fritz 2000, 2002). Their contribution to the
fluctuation amplitude of interest ($P_0$) can be described as
\begin{equation}
P_{\rm fluc} \,=\, P_0 - P_r\,,  \label{pfluc}
\end{equation}
with $P_r$ being the residual fluctuation signal of \textit{undetected}
point sources of faint GCs and background galaxies. Usually the globular
cluster luminosity function (GCLF) is assumed to be Gaussian and the
background galaxy luminosity function to be a power law. Fortunately, even
large uncertainties in $P_r$ ($\sim$20\%) contribute only marginally to the
final error in $\overline{m}$.

In the next step, the total fluctuation amplitude $P_0$ is derived.
The power spectrum of the masked data $P(k)$ consists of a constant $P_0$,
multiplied by the power spectrum of the PSF $E(k)$ and a 
constant $P_1$ (from white-noise component) following 
the relationship (TAL90)
\begin{equation}
P(k) \;=\; P_0\,\times\,E(k) \,+\, P_1\,.  \label{pk}
\end{equation}
It is expected that there is extra noise at very low wave numbers
($k\la25$). However, $P_0$ is tightly constrained by the data across a wide
range of wave numbers and therefore a precise measurement of the fluctuation
flux and variance is possible. We also would like to emphasize that the ratio
$(P_0{-}P_r)/P_1 = P_{\rm fluc}/P_1$ represents another good indicator of the
$S/N$ level.

Finally, the fluctuation magnitude $\overline{m}$ can be derived as 
\begin{equation}
\overline{m}=-2.5 \log(P_{\rm fluc}/T) + m_1.
\end{equation}
Note that by using the masked residual image to create the expectation
power spectrum of the PSF, $E(k)$, the contributions of the target of interest,
background galaxies, GCs, and dust are all excluded, leaving the
fluctuation variance $\sigma^{2}_{f}$ dependent on the mean flux per star
$\overline{f}$ and the mean galaxy flux per pixel $N\bar{f}$
(see equation~\ref{fucflux}).

\section{Observations}\label{obs}

\subsection{Optical SBF}\label{cal}

Since the first successful prescription and quantification of SBFs as a
distance indicator by Tonry \& Schneider (1988), a multiplicity of works have
used this technique to determine distances to globular clusters and
(groups of) galaxies in the optical wavelength regime out to 40 Mpc. After 
this theoretical interpretation, the method was endorsed observationally
with the application to Virgo galaxies in the $VRI$ bandpasses \newline (TAL90).
Tonry (1991) established the first empirical
calibration of the absolute fluctuation magnitude $\overline M_{I}$ with
a zero-point estimate relying solely on the M31 and M32 galaxies. A new
indicator for measuring extragalactic distances was born.

In the following years the SBF method was applied to globular clusters
(Ajhar \& Tonry 1994) and tested in the blue wavelength regime 
(Shopbell et al. 1993; Simard \& Pritchet 1994; Sodemann \& Thomsen 1996).
While the former approach gave good agreement with independent Cepheid
measurements, the latter application yielded a rather poor characterisation
($\overline{m}_B-\overline{m}_V\approx2.45\pm0.15$), mainly due to the
problematic different sensitivities of $\overline M_{B}$ and $\overline M_{V}$
to age and metallicity as well as additional observational limits (e.g., much
fainter fluctuation magnitudes, lower $S/N$, stronger residual signal from
isophotal twist and morphological distortions, dust or disk/ring-like
features, etc).

Several other studies utilised the method for individual galaxies and 
concentrated in refining the SBF method in the red optical bandpasses
(Lorenz et al. 1993; Sodemann \& Thomsen 1995; Fritz 2000, 2002).

John Tonry and collaborators have conducted a
large $I$-band SBF survey to measure the distance to $\sim$300 nearby
galaxies (Tonry et al. 1997, 2000, 2001 [hereafter T01]).
The main result of this programme was an accurate empirical
calibration of the SBF magnitude using optical colours
as (Tonry et al. 2000):
\begin{equation}
\overline M_{I}=(-1.74\pm0.07)+(4.5\pm0.25)\times[(V-I)_0-1.15].
\label{T01} 
\end{equation}
This calibration is based on 
10 Cepheid and 44 SBF distances in 7 different galaxy groups and defined for
galaxies in a colour range of $0.95\leq(V-I)_0\leq 1.30$. 
Both the zero-point and the slope of the relation are well defined (see
Section~\ref{biases} for a detailed discussion of uncertainties in the
calibration of the SBF method). From ground-based $I$-band SBF measurements in
spiral bulges, the zero-point in the $\overline M_{I}$ relation is tied directly
to the Cepheid distance scale to an accuracy of 0.08~mag (excluding
systematic uncertainties in the Large Magellanic Cloud (LMC) distance modulus
with the $P-L$ relationship of $\pm$0.03~mag (Freedman \& Madore 2010) or other
methods, $\pm$0.16~mag (Mould et al. 2000); see Section~\ref{biases}),
corresponding to $\sim$4\% in distance (Tonry et al. 2000).
A comprehensive review of the $I$-band SBF distance survey can be found in
Tonry et al. (1997, 2000) and Blakeslee et al. (1999).

Jensen et al. (2003) used the {\em Hubble Space telescope (HST)}
Near Infrared Camera and Multi-Object Spectrometer
(NICMOS)\footnote[2]{HST/NICMOS:~http://www.stsci.edu/hst/nicmos}
F160W distances to 65 galaxies
(see further Section~\ref{hst}) and established a slightly steeper slope for the
$\overline M_{I}$ relation in the $H$-band than in the $I$-band. These measurements
were tied to the Cepheid distances of the {\em HST} Key Project (KP) 
by Freedman et al. (2001), neglecting the $-0.2$~mag dex$^{-1}$ metallicity
correction. The need for such a correction was ambiguous at that time and argued
that there appeared to be a better agreement with the SBF zero-point predictions
from simple stellar population (SSP) models when the metallicity correction was
omitted. More recent works provide additional evidence of a Cepheid metallicity
dependence (Sakai et al. 2004; Macri et al. 2006; Scowcroft et al. 2009) and
suggest that metallicity effects cannot be neglected when deriving SBF
distances. Further, there is a better understanding of the stellar population
predictions for NIR SBFs (Raimondo 2009; Gonz{\'a}lez-L{\'o}pezlira et al. 2010;
Lee et al. 2010), which is discussed in detail in Section~\ref{mod}.

With respect to the NIR calibration and the collection of high-quality 
{\em HST} Advanced Camera for Surveys
(ACS)\footnote[3]{HST/ACS:~http://www.stsci.edu/hst/acs} data
(see further Section~\ref{hst}),
in recent years there has been some confusion as to whether the T01 distances
derived from of the ground-based $I$-band SBF distance survey (Tonry et al.
1997) require some correction.  
To overcome this discrepancy, Blakeslee et al. (2010) suggest to apply a
correction to distances determined using the relation in equation~\ref{T01},
including both a zero-point (resulting from a revised Cepheid distance
calibration of the KP (Freedman et al. 2001), which is $-0.06$~mag for the
T01 distance moduli)
and a second-order bias correction (originating from a distance bias in 
low-quality data of T01). Depending on the quality of the T01 data, the
correction in the final T01 distance moduli $(m-M)_{\rm T01}$ 
ranges between $+0.05$~mag (poorest quality) to $\sim-0.12$~mag (highest
quality). In order to be consistent with the revised Cepheid scale, the corrected
T01 calibration, and other works using a metallicity correction, distance moduli
obtained with the Jensen et al. (2003) IR SBF calibration would need to be
increased systematically by $+0.10$~mag.

Ground-based optical SBF observations are restri\-cted to relatively low redshifts.
Therefore, these measurements are affected by the local peculiar velocities and
difficulties are experienced in deriving direct numbers for the Hubble constant
$H_0$. In combination with other distance estimators (e.g., $D_n-\sigma$,
Tully-Fisher relation, see Section~\ref{hstopt} for details),
or space-based measurements, the SBF method provides much tighter
constraints on the Hubble flow, as more distant galaxies can be considered.
Using Fundamental Plane distances to $\sim$170 elliptical galaxies,
Blakeslee et al. (2002) derived
$H_0=72\pm4~({\rm random})\pm11~({\rm systematic})$ \kms Mpc$^{-1}$.

\subsection{Near-Infrared SBF}

There are several obvious advantages for the application of the SBF technique
at longer wavelengths. The fluctuation signal is significantly stronger,
$\sim$35 times brighter in the $K$ than in the $I$-band, making the SBF signal
detectable to greater distances. The seeing is typically much better in the IR
than in the optical and this greatly improves the SBF amplitude, which is
inversely proportional to the seeing FWHM. As the
fluctuations are red compared to the globular cluster population, the resulting
contrast between stellar and GC contributions to the SBF signal is higher and
can be easier separated. A drawback is that the sky background is larger in
$K$. However, because of the red fluctuations, higher $S/N$ ratios can be
reached in shorter exposure times than in the optical. Further, the extinction
correction is a factor of 4 lower in $K'$ than in $I$ and for old, metal-rich
populations the stellar population models predict a weaker dependence of the
$K$-band SBF on the integrated $(V-I)$ colour than in the $I$-band, reducing
uncertainties in the colour determination of the stellar system (Worthey 1993).
For this reason, the effects of age and metallicity are largely not degenerate.
This fact acts in favour for stellar population studies of galaxies, but is
less beneficial for simple distance calibrations.

The first feasibility studies of the IR SBF method were performed by
Luppino \& Tonry (1993); Pahre \& Mould (1994) and Jensen et al. (1996) on a
small set of galaxies in the Local Group and the Virgo Cluster. These works
reported similar $\overline M_{K}$ calibrations of
$\overline M_{K}\sim-5.65\pm0.20$~mag. Jensen et al. (1998) extended the IR
calibration to nine galaxies in the Fornax and Eridanus clusters using a wider
range of $(V-I)$ colours, metallicities and additional Cepheid distance
measurements. Using high $S/N$ ground-based NIR observations,
Jensen et al. (1999) pushed their ground-based $K$-band observations to measure 
distances to elliptical galaxies in the Hydra ($cz=4054$ \kms) and the
Coma ($cz=7186$ \kms) clusters. The authors suggested that $\overline{M}_K$
itself can be used as a standard candle and therefore give direct constraints
on the Hubble constant $H_0$ (see Section~\ref{dis} for details). Based on the
Coma SBF distance, this work derives $H_0=85\pm11$ \kmsM.

\subsection{SBF from {\em HST}}\label{hst}

A number of works have mined the {\em HST} archive to use galaxy observations of other
programmes for SBF analyses. After the {\em HST} refurbishment in December 1993
(STS-61, HST-SM1), SBF observations have been collected from several allocated
programmes through Cyles 5 to 7 and 15 using the Wide Field Planetary Camera 2
(WFPC2)\footnote[4]{HST/WFPC2:~http://www.stsci.edu/hst/wfpc2}, ACS
and NICMOS as well as a WFPC2 Guaranteed Time
Observation (GTO) programme (Pahre et al. 1999). Further, new Wide Field
Camera 3 (WFC3/IR)\footnote[5]{HST/WFC3:~http://www.stsci.edu/hst/wfc3}
observations are executed during Cyle 17.

\subsubsection{Optical}\label{hstopt}

Recent work on SBF distance determinations has concentrated on
early-type galaxies observed with the {\em HST}. Space-based
observations offer a number of advantages over ground-based measurements:
Higher spatial resolution (FWHM$_{\rm WFPC2}=0.18''$,
FWHM$_{\rm ACS}=0.098''$), easier identification of GCs and dust due
to the higher spatial resolution (reaching $\sim$2 mag fainter systems),
higher PSF stability and photometric calibration,
lower sky background, no disturbing atmospheric absorption hence no
extinction correction, and insensitivity to focus changes of the camera
(as in the Fourier analysis only wave numbers outside the $\overline{m}$
range are affected, see Section~\ref{pract} for details).

Using the WFPC2 onboard the 
{\em HST}, Ajhar et al. (1997) targeted 16 galaxies in the F555W ($V$)
and F814W ($I$) filters. They found a steeper calibration of
$\overline M_{I,F814W}=(-1.73\pm0.07)+(6.5\pm0.7)[(V-I)_0-1.15]$
compared to Tonry et al. (1997). However, the usage of the WFPC2 camera and the
harder calibration of the broader F814W was impractical for measuring distances
greater than the Coma cluster (104.7~Mpc) because the observing time increases
with the square of the distance.

Based on WFPC2 observations from a dedicated IDT/GTO programme, Pahre et al. (1999)
have investigated the distances to NGC~3379, NGC~4373, and NGC~4406. One
important goal of this work was to demonstrate the capability of the WFPC2
camera (despite its sampling rate) to measure extragalactic distances using the
SBF technique. The refurbishment was a success and the results for the three
galaxies were in good agreement with previous measurements. For the bright
group galaxy NGC~4373 in the Hydra-Centaurus supercluster, the authors derive a
distance of $d=39.6\pm2.2$~Mpc and a peculiar velocity for the galaxy of
$415\pm300$ \kms. This peculiar velocity is about half as large as the
$838$ \kms peculiar velocity prediction using the Great Attractor model from
Lynden-Bell et al. (1988) based on $D_n-\sigma$ observations.

Ajhar et al. (2001) measured the SBF signal to E+S0s that have Type Ia 
Supernovae (SNe Ia) and found no systematic error with distance. However, they
detected a systematic offset of $\sim$0.25 mag because of different Cepheid
calibrations.

The higher quantum efficiency (a factor of 2 to 10, depending on the wavelength,
see Table~\ref{qe}) and larger field of the Advanced Camera for Surveys Wide
Field Channel 
(ACS/WFC)\footnote[6]{ACS-WFC:~www.stsci.edu/hst/acs/Detector/detectors}
compared to the WFPC2 chip 
($202''\times202''$ vs. $134''\times134''$) and better sampling of the PSF
(factor of $\sim$2) allows researchers to extend the SBF distance scale
to distances greater than 10,000 \kms.
In an effort to study an unbiased, almost complete sample of nearby galaxies,
Tonry and collaborators have obtained WFPC2 and ACS observations to
$\sim$300 galaxies (Tonry et al. 2000, T01). 
The $I$-band SBF survey provided accurate calibrations 
and measurements to 91 elliptical and lenticular (E+S0) galaxies, and 9 spiral
bulges up to 2000 \kms, thereby building a link to
other distance indicators that are sensitive to and extend well out into the
Hubble flow, e.g., $D_n-\sigma$ (Dressler et al. 1987; Lynden-Bell et al. 1988),
Tully-Fisher relation (Aaronson et al. 1989; Giovanelli et al. 1997), or
SNe Ia (Sandage \& Tammann 1982; Riess et al. 2009).

\begin{table}[t]
\begin{center}
\caption{Comparison of the Filter Efficiency between WFPC2 and ACS Filters.}
\label{qe}
\medskip
\begin{tabular}{ccc}\\
\hline 
Filter  &  WFPC2  &   ACS  \\
        & \%      &    \%    \\
\hline 
F450W$^a$  ($U$)  &  8.5  &  36  \\
F555W  \   ($B$)  &   11  &  37  \\
F606W  \   ($V$)  &   14  &  44  \\
F702W$^b$  ($R$)  &   14  &  42  \\
F814W  \   ($I$)  &   10  &  42  \\
F850LP \   ($z$)  &  3.9  &  25  \\
\hline
\end{tabular}
\medskip\\
$^a$For ACS the values for F475W were adopted.\\
$^b$For ACS the values for F775W were adopted.\\
Numbers are based on the WFPC2 instrument handbook\footnotemark[7].
\end{center}
\end{table}
\footnotetext[7]{HST-IHB:~http://www.stsci.edu/hst/wfpc2/documents/\newline handbook/IHB\_17.html}

More recently the SBF technique was applied to derive distances to dwarf
galaxies in groups and clusters in the southern hemisphere using large
ground-based telescope facilities (e.g., Jerjen et al. 2001, 2004;
Mieske et al. 2006, 2007; Dunn \& Jerjen 2006; Mei et al. 2007;
Blakeslee et al. 2009; 2010) or using {\em HST} ACS/WFC observations to nearby
early-type galaxies (Cantiello et al. 2005, 2007a,b; 
Barber DeGraaff et al. 2007; Biscardi et al. 2008).

The first optical SBF distance out to $\sim$100 Mpc using ACS/WFC F814W bandpass
observations to shell galaxies was measured by Biscardi et al. (2008).
Barber DeGraaff et al. (2007) demonstrated that SBFs are also a powerful
tool to determine the distance to barred lenticular galaxies. For the nearby
SB0 galaxy NGC~1533 in the Dorado group, the authors derive
$(m-M)=31.44\pm0.12$~mag, corresponding to $d=19.4\pm1.1$~Mpc.
More recently Blakeslee et al. (2010) established an empirical calibration of
the SBF method for the ACS/F814W filter (similar to the Johnson $I$-band).
Because of the much higher throughput, the F814W ($I_{814}$) bandpass is more
efficient and therefore to be preferred over the ACS/F850LP ($z_{850}$)
bandpass (see Table~\ref{qe} for a comparison).
However, the most extensive observing campaign has been performed in the
ACS/WFC $z_{850}$ band as part of the ACS Virgo and Fornax Cluster
Surveys (ACSVCS, ACSFCS). In total, distances to 135 nearby galaxies were
derived, for 90 early-type galaxies in the Virgo cluster
(Mei et al. 2005, 2007) and 43 galaxies in the Fornax cluster
(Blakeslee et al. 2009); the latter also give a refinement of the calibration
of the combined cluster samples. Moreover, the SBF distances to two dwarf
galaxies in Virgo were presented in Blakeslee et al. (2010).

\subsubsection{Near-Infrared}

Using {\em HST}/WFPC2 observations to the galaxy NGC 4373 in the
Hydra-Centaurus supercluster, Pahre et al. (1999)
found a good agreement (difference in $P_0\sim2$\%) with the results by
Tonry et al. (1990, 1997), the latter using a different reduction and
analysis package.

Jensen and collaborators measured the IR SBF distances to 16 galaxies
located in the Leo, Virgo and Fornax clusters
with $\leq$10,000 \kms using the F160W NICMOS camera onboard the {\em HST}
(Jensen et al. 2001). They were not able to give tight constraints on the
Hubble constant. For the Hubble constant they derived 
$H_0=72\pm2.3-76\pm1.3({\rm statistic})\pm6({\rm systematic})$ \kmsM, depending
on the distance of the galaxies under consideration. Because the Local Group is
located in an underdense region of the universe, the measurement is biased
and therefore results in uncertainties and in a larger value of $H_0=76$~\kmsM. 
Soon thereafter, the team extended their efforts and established
{\em HST}/NICMOS F160W SBF measurements for 65 E+S0
galaxies located in different environments (Jensen et al. 2003). Interestingly,
they found evidence for intermediate age stars in the stellar populations of
the galaxies. The central bluer colours of the E+S0 galaxies suggested younger
stellar populations showing signs of recent star formation that are more metal
rich. The SBF technique is therefore an independent tool to give insights to
the composite stellar populations of early-type galaxies.

%
%
\begin{figure}[!t]
\begin{center}
\includegraphics[angle=0,width=3.0in]{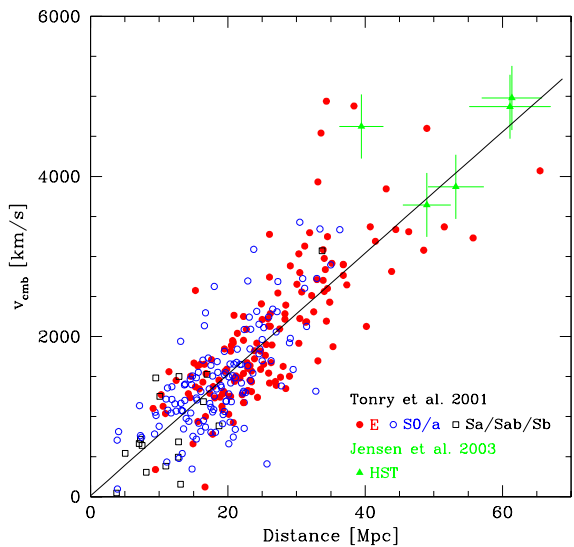}
\vspace{0.2cm}
\caption[SBF gradients]{\small{Hubble diagram for galaxies with SBF distances.
Recession velocities in the CMB frame are plotted as a function of distance for
ground-based data (Tonry et al. 2001, circles and squares) and {\em HST}/NICMOS
measurements (Jensen et al. 2003, triangles). The line has a slope of 75 \kmsM.
Typical 1$\sigma$ uncertainties are indicated for the {\em HST} data. For clarity,
error bars are omitted for the ground-based data. The very deviant points
near 39 Mpc are galaxies with high peculiar velocities belonging to the
Cen 45 cluster of galaxies.}}
  \label{sbfhub}
\end{center}
\end{figure}

 Figure~\ref{sbfhub} shows the Hubble diagram for galaxies that have
SBF distances. The redshift in the cosmic microwave background (CMB) frame
is displayed as a function of distance for ground-based data of the $I$-band
SBF survey, divided into different galaxy types (Tonry et al. 2001),
complemented by additional high-$S/N$ {\em HST}/NICMOS F160W SBF measurements
(Jensen et al. 2003). All redshifts were corrected for the
peculiar velocity of the Local Group of galaxies with respect to the CMB.
The expansion of the universe is clearly apparent and a linear, unconstrained
fit to the data gives a Hubble ratio of 
$75\pm4({\rm stat})\pm7({\rm sys})$~\kmsM. Local peculiar velocities are
the origin for scatter in the Hubble ratio at distances of 40 Mpc, but the
majority of data at this distance are pretty well described by the derived
Hubble value. Part of the scatter in the SBF Hubble diagram is due to
contributions of external sources (e.g., differences in the stellar
populations), but does not solely rely on uncertainties in the distances.
In the next few years, the SBF Hubble diagram will be better defined at
distances reaching out into the Hubble flow ($d>40$ Mpc) and likely be extended
up to 100 Mpc using high-quality measurements of the WFC3/IR instrument onboard
the {\em HST}.

%
%
\begin{figure}[!t]
\begin{center}
\includegraphics[angle=0,width=3.0in]{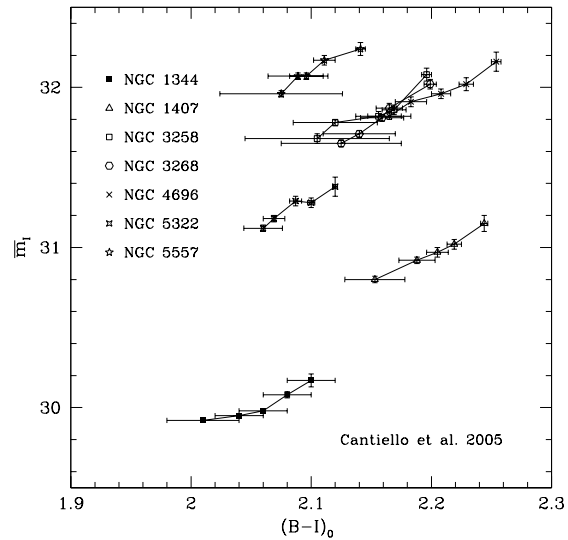}
\vspace{0.2cm}
\caption[SBF gradients]{\small{Apparent $I$-band SBF magnitudes in a series of
concentric annuli plotted as a function of $(B-I)_0$ galaxy colour
(Cantiello et al. 2005). With increasing radial distance from the galaxy centre,
the mean galaxy colour becomes bluer and $\overline{m}_I$ brighter. The vertical
offsets are due to the different distances of the galaxies.}}
  \label{sbfg}
\end{center}
\end{figure}

\subsection{Stellar Population Gradients}\label{grad}

Early-type galaxies are well known to show radial colour gradients  due to
variations in their stellar population properties (e.g., de Vaucouleurs 1961;
Peletier et al. 1990). The first SBF gradients were discovered in
ground-based observations for individual galaxies (Ton\-ry 1991;
Sodemann \& Thomsen 1995, 1996; Jensen et al. 1996; Fritz 2000, 2002).

Using ACS/WFC observations, Cantiello and collaborators have
investigated the SBF gradients of a larger sample of nearby galaxies
(Cantiello et al. 2005, 2007a). Examples of SBF colour
gradients are shown in Figure~\ref{sbfg}. The size of the slope between the
intrinsic internal galaxy SBFs ($\overline{M}_I$) and the galaxy colour
($(B{-}I)_0$) for multiple annuli within a galaxy allows one to disentangle
whether the gradients are due to age or metallicity variations. For the
majority of galaxies in Figure~\ref{sbfg}, the gradients are metallicity driven
(e.g., NGC\,3258), whereas for a few cases the detection of a shallower slope
suggests an age gradient which is equally or more important (e.g., NGC\,1344).
Galaxies with age gradients indicate a bluer, younger stellar
population at larger galaxy radii and some evidence for recent merging 
activity or accretion events. More studies on this topic involving larger
samples would allow scientists to draw further conclusions on the origin of
SBF gradients.

For the foreseeable future, the WFC3/IR camera onboard {\em HST} offers a powerful
new possibility of investigating the internal SBF gradients in the NIR (see also
Section~\ref{fut}). It would be extremely exciting to compare the optical
gradients with their NIR counterparts. In the NIR, pure age or metallicity
variations affect the SBF magnitude dependence on galaxy colour very 
differently. More comprehensive studies on SBF gradients, in particular in
the NIR, might allow us to disentangle the age/metallicity degeneracy in
elliptical galaxies and therefore provide new insights into their different
evolutionary histories.

\section{SBF as a Standard Candle}\label{dis}

To derive accurate distances with the SBF technique, it is necessary that the
pixel-to-pixel fluctuations are dominated by the Poisson statistics of the
stars rather than by the photon shot-noise statistics or other detector
artefacts (Tonry \& Schneider 1988). To fulfil this constraint, one needs to
acquire high $S/N$ images (see Section~\ref{pract}).
Furthermore, other fluctuations that originate from unwanted contributions
($P_r$) such as GCs, foreground and/or background galaxies, need to be
removed as well (Blakeslee \& Tonry 1995; Jensen et al. 1999). In particular,
for distance moduli $(m-M)>32$, the fluctuations from external
sources are at least of the same order of magnitude as (or even higher than) the
SBF fluctuations themselves and therefore need to be carefully removed before
estimating distances (Jensen et al. 1998). This is even valid for the NIR bands,
although one might expect that the relative contributions to the fluctuations
from GCs to be smaller, as the stellar SBFs are much redder than (those of) GCs.
Since the background is higher in the NIR and the sampling and sensitivity to
faint point sources is decreased, this effect cancels out for the total
fluctuation power (contribution of background galaxies is reduced by a factor
of $\sim$10).

There are two main requirements for using the SBF technique as a distance
indicator: {\it (i)} The bright end of the LF among galaxies of different type
(ellipticals, S0s and spiral bulges) is universal, or {\it (ii)} The variations in
the LF from stellar system to system can be detected and corrected for, so
that $\overline{M}$ remains calibrated at high precision, hence a standard
candle (see also Section~\ref{met} and equation~\ref{LF}).

Several studies have proven that absolute fluctuation magnitudes in both the
$I$ and the $K$-band of elliptical galaxies are correlated with the internal
$(V-I)_0$ galaxy colour (TAL90; Luppino \& Tonry 1993; Ajhar et al. 1997;
Tonry et al. 1997; Jensen et al. 1998; Mei et al. 2001; Liu et al. 2000, 2002).
As a consequence the empirical calibration can be considered as universal
and therefore as a standard candle. The linear relationship between absolute
fluctuation magnitude and galaxy colour has been calibrated to Cepheid distance
measurements in the Virgo Cluster and the Local Group (see Section~\ref{cal}
above). The intrinsic scatter in the Cepheid zero-point calibration is 
$\sim$0.05-0.09 mag, whereas the stellar population models (Worthey 1993)
suggest an uncertainty in the $\overline{M}_I-(V-I)_0$ relation of $<0.11$ mag.
The exact scatter in the theory depends on the composition and the
variations among the stellar populations.

Jensen et al. (1999) compared their $K$-band data to the theoretical
predictions by Worthey (1993) and concluded that their fluctuation magnitudes
$\overline{M}_K$ are almost constant in the colour range
$1.10\leq(V-I)_0\leq 1.30$. This implies that the $K$-band SBF itself is a
good standard candle with a median absolute magnitude of
$\overline{M}_K=-5.57\pm0.19$. The average scatter in $\overline{M}_K$ is only
0.06 mag, but the models with solar metallicity suggest brighter
absolute magnitudes of 
$\overline{M}_{Ks}=-5.84\pm0.04$, but a slope ($3.6\pm0.8$)
which is in agreement with the optical calibration (Liu et al. 2002).
From this comparison it is clear that more high-quality observational data
are needed and a more careful description of the stellar populations has to be
obtained from a theoretical point of view. A more detailed discussion on
this issue with the presentation of new stellar population models is given in
Section~\ref{mod}.

\subsection{Caveats for Distance \newline Determinations}\label{biases}

As we have shown, SBF magnitudes represent a universal standard candle, and
hence a precise distance indicator. However, there are several possible
problems and uncertainties that could affect accurate distance measurements.

\subsubsection{Target Selection}

The preferred targets for SBF measurements are dynamically hot stellar
systems, either GCs or elliptical galaxies. There have been attempts to
measure distances to disky S0 galaxies (NGC~1375: Lorenz et al. 1993)
or edge-on spiral bulges (NGC~4565: Simard \& Pritchet 1994; NGC~3115: T01).
However, apart from differences in the analysis techniques, Cepheid stars are
only found in the dusty disks of spiral galaxies. Stellar systems with multiple
components in the light profile (bulge, disk, bar, rings) are difficult to model
for the SBF method, as the stars of the disk need to be correlated to the place
where the SBF signal was measured. Hence, the SBF technique is mostly limited
to the dust-free stellar systems of elliptical galaxies.
An alternative approach is to go to space. Thanks to the superb resolution of
{\em HST} compared to ground-based telescopes, different components and dust
features can be detected, modelled, and subtracted much better and the SBF
technique extended to S0 galaxies and early-type spiral bulges
(Jensen et al. 2003). This approach appears to be promising, as Cepheids are
most common and have a higher abundance in the outer disks of spiral galaxies,
which are usually difficult to analyse with the SBF technique.

\subsubsection{PSF Modelling}

A precise stellar template is crucial for measuring the exact SBF amplitude.
All modifications of the power spectrum of the stellar template translate
directly to the power spectrum of the fluctuations. For example, a
mismatch of 3\% in the normalisation of the stellar power spectrum 
results in an error of 0.03 mag in the fluctuation magnitude
$\overline{m}$. Usually, having a good stellar template as PSF is not a major
problem, as there are stars located close to or in the surrounding field of the
target galaxy. In the NIR, however, this issue is more prominent as the stars
must be detectable and separated from the much higher sky background. For
{\em HST} there is the possibility of constructing synthetic PSF templates to
account for the PSF variations and distortion effects on the pixel-sensitive
WFPC2, ACS or NICMOS chips (e.g., Fritz et al. 2009a,b). In general, for
ground-based observations the mismatch between real and synthetic PSFs is small
(see Fritz 2002), but when real observed PSF are available these are to be
preferred.

At the present time, the real limiting factor of (N)IR detectors is the small
field-of-view relative to their optical counterparts. As a consequence, fewer
bright field stars are available for calibration purposes that are
uncontaminated by galaxy or companion light.

%
%
\begin{figure}[!t]
\begin{center}
\includegraphics[angle=0,width=3.0in]{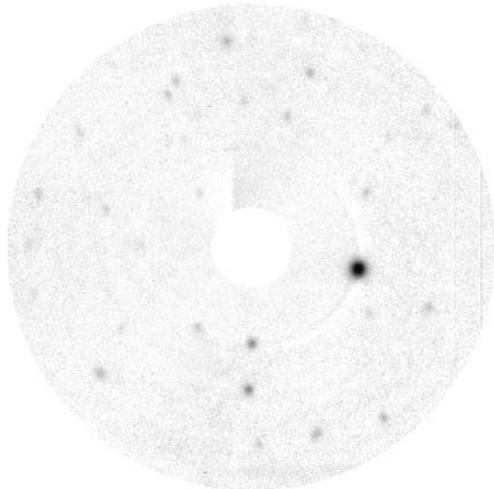}
\vspace{0.2cm}
\caption[SBF sources]{\small{Masked (galaxy-subtracted) residual image of
NGC~3379 (adapted from the investigation of Fritz 2002). The unsaturated bright
star right from the centre of the galaxy was adopted as a PSF
template. Sources in this image consist of GCs, background galaxies
and SBFs. From the normalized luminosity fluctuations,
a characteristic flux-weighted mean flux of $\bar{f}=3.37$
ADU/star/300s was measured, which corresponds to a weighted mean
fluctuation magnitude of $\overline{m}=28.61\pm0.04$~mag (Fritz 2002).}}
 \label{bgrs}
\end{center}
\end{figure}

\subsubsection{Background Sources}

Gound-based SBF measurements have a crucial limiting factor: The detection,
modelling and careful removal of faint background sources and globular clusters.
The total fluctuation amplitude ($P_0$) must be dominated by the SBF
signal, hence the residual fluctuation of {\em undetected} faint GCs and
background galaxies ($P_r$) needs to be smaller than the SBF fluctuations.
This is illustrated in Figure~\ref{bgrs}.
The superior resolution of space-based observations offers the great advantage
of much more accurate detection and sensitivity to unwanted sources. Further, in
the IR the sky background of NICMOS {\em HST} observations is more than a
factor of 100 lower compared to IR observations from ground-based telescopes.

\subsubsection{The Good, the Bad, and the Ugly}

Considering the challenge involved in deriving a reliable estimate of the
fluctuation magnitude $\overline{m}$, the measurement of the galaxy colour
should represent an easy task. Instead, however, the linear correlation of
$\overline{M}_{I}\propto 4.5(V-I)$ makes an accurate colour determination
essential. Internal colour gradients in a galaxy introduce complexity and
therefore cause difficulties when deriving the colour of a stellar system.
Both $\overline{m}$ and $(V-I)$ need to be measured over the same region in
the galaxy, which is tricky if the galaxy has multiple components of bulge
plus disk and $\overline{m}$ is established in the centre, whereas $(V-I)$
is obtained from the outer regions of the disk. 

For optical colours the effects of age and metallicity are largely
degenerate. Therefore, going to the IR and looking for a NIR colour calibration
would be greatly beneficial as it would allow one to quantify differences in
the stellar populations that alter the light-averaged age and metallicity of a
stellar system. A few observational (Jensen et al. 2003) and theoretical
approaches (Blakeslee et al. 2001a; Lee et al. 2010) in this direction have
been conducted. However, more observations and improved model descriptions
would finally provide an accurate handle on the age/metallicity degeneracy.

Moreover, SBF magnitudes are required to be corrected for the dust extinction
caused by our own Milky Way galaxy. Usually, the 100$\mu$m DIRBE/{\em IRAS}
dust emission maps by Schlegel, Finkbeiner \& Davis (1998) are adopted. In
the future, improved galactic extinction values could be derived from
multi-wavelength studies involving mid-IR or {\em Spitzer}/IRAC combined with
blue ultraviolet (UV) {\em GALEX} observations.

The uncertainty in the zero-point of the Cepheid calibration is less than 0.1~mag
(see Section~\ref{dis} above). However, the revised Cepheid calibration
(Freedman et al. 2001) depends on the metallicity of the underlying stellar
population and needs to be accounted for (Sakai et al. 2004; Macri et al. 2006;
Scowcroft et al. 2009). This additional metallicity correction is of the order
of $-0.06$~mag to the distance moduli and therefore adds only a small
uncertainty of $0.08-0.10$~mag when measuring distances based on SBFs
(Jensen et al. 2003; Blakeslee et al. 2010; see also Section~\ref{cal}).

\section{Theory meets Observation}\label{mod}

\subsection{Origin of SBFs}

\begin{figure}[!t]
\begin{center}
\includegraphics[angle=0,width=3.0in]{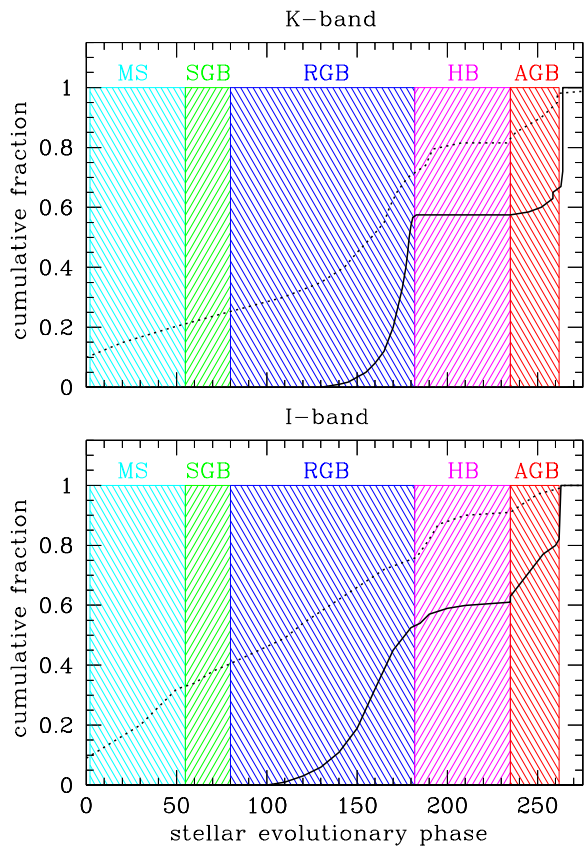}
\vspace{0.2cm}
\caption[SBF origin]{\small{
Origin of the SBF signal. For an old, solar-metallicity, single-burst stellar
population model (see text) the fraction of the total light is given as a
function of stellar evolutionary phase.
Evolutionary phases range from the ZAMS (left) to the late
AGB phase (right) in arbitrary units. The integrated flux (dotted line) arises
from all stars of all stellar evolutionary phases. In contrast, the IR SBF
signal (solid line, top panel) originates entirely from the RGB and AGB phase,
with some additional contribution from the HB for optical SBFs (solid line,
lower panel).}}
  \label{sbfo}
\end{center}
\end{figure}

Since the absolute SBF magnitude $\overline{M}$ in a given wavelength range
depends on the underlying stellar population properties, SBF luminosities in
different bandpasses probe different stages of the stellar
evolution within the unresolved stellar host system.
UV and blue SBF magnitudes are sensitive to the evolution of stars within the 
hot Horizontal Branch (HB) and post-Asymptotic Giant Branch (post-AGB) phase;
optical SBF magnitudes serve as a measure of stars within the RGB and the HB
range, whereas NIR SBF luminosities represent a stellar evolution stage within
the AGB and the TP-AGB phase. Since the SBF luminosity is weighted by the
square of the stellar luminosity (see equation~\ref{LF}), it is extremely
sensitive to the most luminous giant stars. Figure~\ref{sbfo} illustrates the
higher sensitivity of the SBF signal to luminous cool giant stars in their late
stellar evolutionary stages compared to the total integrated stellar flux
signal. The stellar evolutionary phases are ordered according to their relative
contribution to the total integrated light and range from the Zero-Age
Main-Sequence (ZAMS) to the late AGB phase. For the computation a 12-Gyr old,
solar-metallicity, single-burst stellar population model using the Padova
evolutionary tracks with semi-empirical spectral energy distributions (SEDs)
was adopted (Mar\-igo et al. 2008).
It can be seen that the integrated flux arises from stars throughout all stellar
evolutionary phases. In contrast, the IR SBF signal originates entirely from
the RGB and AGB phase ($\sim$88\% of the flux is within the two brightest
IR magnitudes), whereas for optical SBFs some additional contribution comes
from stars along the HB ($\sim$60\% of the signal is concentrated within
the 3-4 brightest optical magnitudes).

%
%

From Figure~\ref{sbfo} it becomes evident that the SBF luminosities and colours
should closely trace the peak luminosity and colours of late cool stars along
the giant branch. Therefore, in the following section we give a brief summary
of the main evolutionary phases of cool giant stars, the RGB phase (H-shell
burning) and the AGB phase (He-shell burning). 

\subsection{The Late Stages of Cool Giant Stars}\label{agb}

Low- and intermediate-mass stars of the main-sequence ($1\leq M/M_{\odot}\leq 8$)
experience high luminosities as red giants in their late evolutionary phase
along the AGB (up to a few $10^{4} L/L_{\odot}$). However, due to their low
effective temperatures ($T_{\rm eff}\leq3500$~K), these AGB stars are highly
extended objects with radii and cool atmospheres of several $100~R/R_{\odot}$.
Depending on the initial ZAMS mass of the star, the
hydrogen (H) in the stellar core will be exhausted, and subsequently the Helium
(He) core will rapidly contract and heaten up, while the H will continue to
burn in a shell around the nucleus.
This stage marks the start of the RGB phase (H-shell burning).
The beginning of the AGB phase is characterised by the end of He-burning in
the stellar core. In the first phase, the He-burning continues in a shell
around the core (early AGB). On the upper part of the AGB, the star becomes
unstable because of strong radial pulsations with typical time scales of a
few tens to hundreds of days (e.g., Iben \& Renzini 1983;
Vassiliadis \& Wood 1993; Dorfi \& Hoefner 1998). The star enters the TP-AGB
phase of alternate H- and He-shell burning. In this phase, the AGB
star shows thermal pulses: The H-shell burning gets interrupted at regular times
by the ignition of the He that has accumulated under the H-shell beforehand as
a byproduct of H-burning. Further, the variable convection zones mix material
processed in the nucleus to the stellar surface, thereby changing the chemical
composition of the convective layers, i.e. dredge-up phases
(Iben \& Renzini 1983). Heavier elements (mainly carbon, nitrogen, oxygen, and
s-process elements) are ultimately transported outwards into the stellar
photosphere and modify the stellar surface composition.
The relative fraction of carbon-to-oxygen ([C/O]-ratio) is defined by the core
mass and this mixing ratio regulates the generation of different molecules in
the cool stellar atmosphere (e.g., TiO, C2, CN). Most frequent are
oxygen-rich AGB stars (M-star: [C/O]$<$0.95), but sometimes a carbon-rich star
([C/O]$>$1.0) is formed, which eventually develops into a central white dwarf
star within a planetary nebula.

Since the SBF signal is more sensitive to the most luminous stars in a stellar
system, the SBFs provide tighter constraints on the evolution of evolved
cool giant stars in elliptical galaxies than integrated luminosities alone.
However, several aspects of the physics of cool giant stars (e.g., interior
structure, mass-loss, mixing length for internal convection, pulsation modes),
as well as their exact spectral energy distributions, remain poorly understood.
Ideal observational targets would be nearby GCs, which host these stellar
populations of homogenous composition and age. However, because cool giants
cross their late evolutionary phase rapidly, only a handful of these stars exist
at the present time for a given cluster. To overcome uncertainties in GC
analyses such as small number statistics or cosmic variance, SBF measurements
of individual galaxies can represent an alternative approach, as the SBF signal 
is dominated by much higher luminosities and the stellar systems are expected
to comprise metal-rich populations.

Since a few years, more attention has been dedicated in observing young
intermediate-age stellar populations (0.5$\la$ Gyr\,$\la$3) in GC and
stellar clusters in the MCs (see e.g. Gonz{\'a}lez-L{\'o}pezlira et al. 2005;
Raimondo et al. 2005; Mouhcine et al. 2005; Lee et al. 2010).
The brightest cluster stars are carbon-type AGB stars with bolometric luminosities 
that exceed the brightness of the tip of the RGB. In contrast to M-type AGB
stars, carbon AGB stars are more luminous and also redder.

Therefore, IR SBFs represent a promising tool to disentangle the
effects and contributions of ages and metallicities in unresolved stellar
systems. In particular, mid-infrared (MIR) SBFs in the $L$ (3.5~$\mu$m) and
$M$-band (4.8~$\mu$m) are expected to be highly sensitive tracers of the
age of a stellar population.
Because of the extremely high thermal IR background of the earth's
atmosphere, however, observations from the ground are only possible for the
very nearest and bright galaxies (e.g., M31). In the foreseeable future, a new
window on MIR SBFs will be opened with the {\em James Webb Space Telescope
(JWST)}, which will access both the $L$- and $M$-bands.

%
%
\begin{figure*}[t]
\begin{center}
\includegraphics[angle=0,width=3.20in]{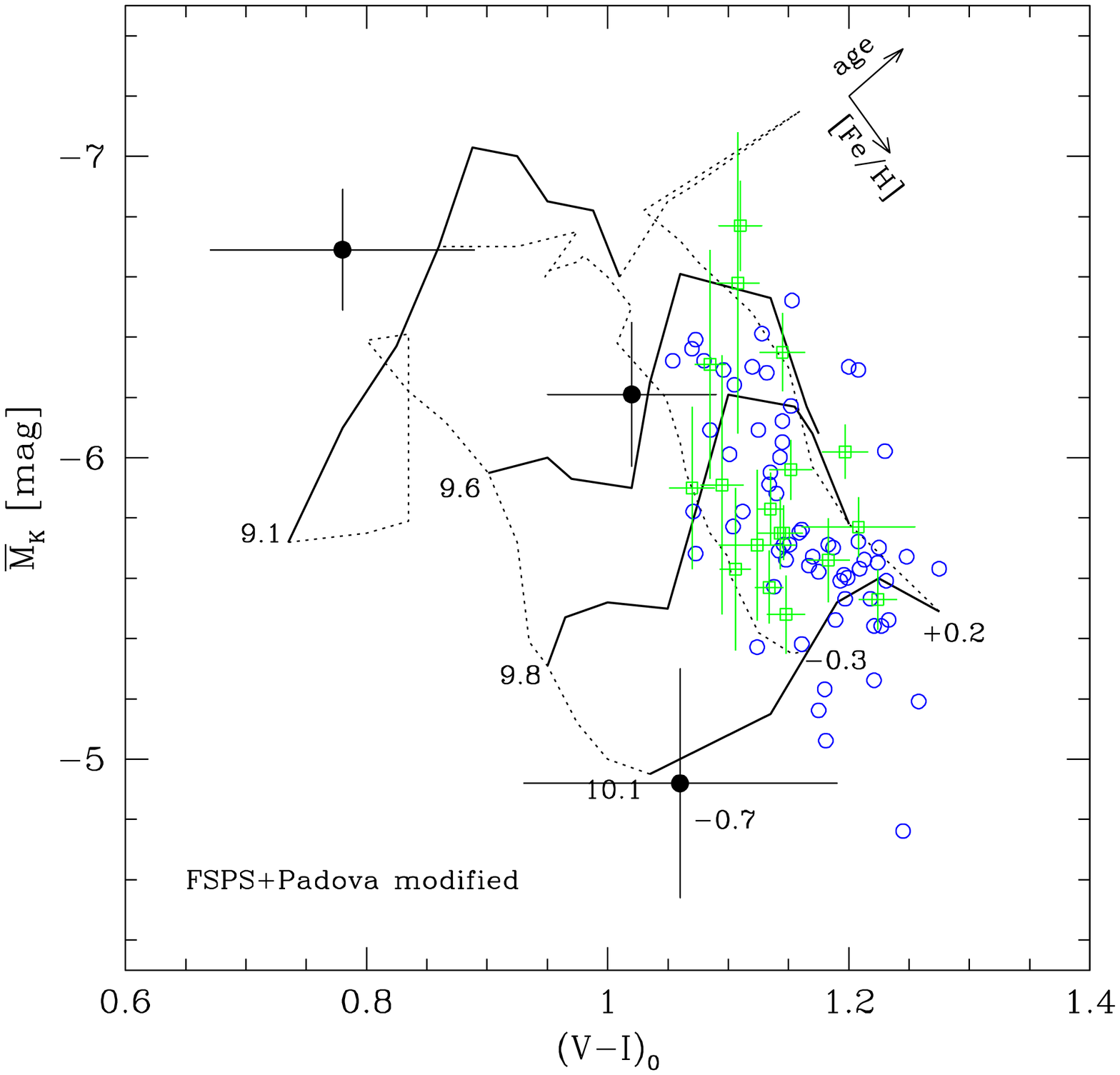}\nobreak
\includegraphics[angle=0,width=3.20in]{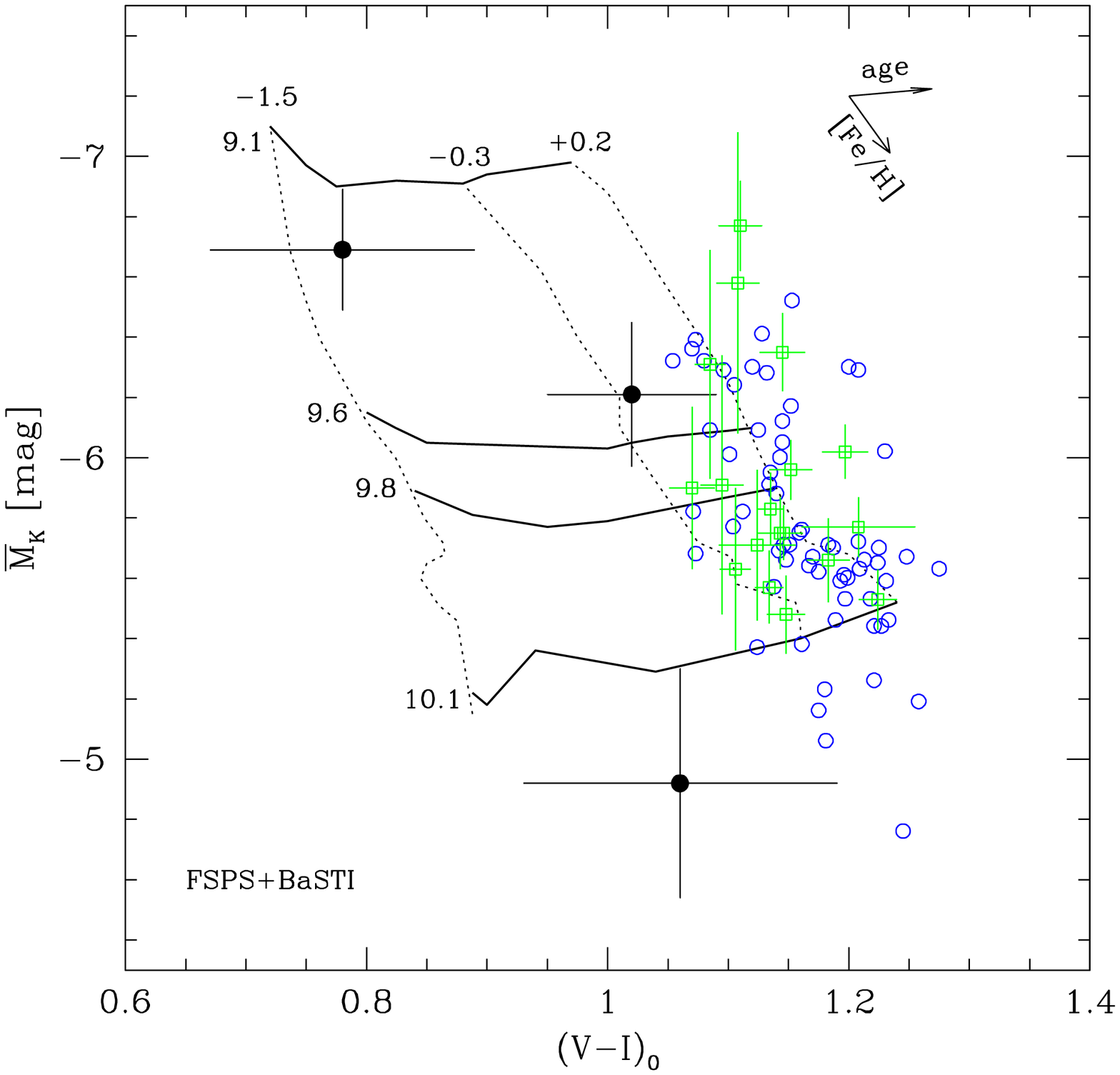}
\vspace{0.2cm}
\caption[SSP models]{\small{$K$-band SBF fluctuation magnitude measurements
and FSPS models as a function of integrated $(V-I)_0$ colour. Different FSPS model
predictions (Conroy \& Gunn 2010, lines) are compared to observations of elliptical
galaxies in the Fornax cluster (Liu et al. 2002, green squares), ellipticals in a
variety of environments (Jensen et al. 2003, blue circles), and star clusters in the
MCs (Gonz{\'a}lez-L{\'o}pezlira et al. 2005, solid black circles). For clarity
reasons, uncertainties are displayed only for the Fornax galaxies and
MC clusters. Model predictions are shown as lines of constant age (solid
lines; in units of log(t/yrs)) and lines of constant metallicity (dashed lines;
in units of log(Z/Z$_{\odot}$)). Arrows denote the direction of 0.02 and
0.05~dex variations in age and metallicity, respectively. FSPS models
are constructed with the modified Padova isochrones (left panel) and the BaSTI
isochrones (right panel). The unmodified Padova isochrones used in the past give
only a poor approximation of half of the data. Large metallicity variations
of $\Delta {\rm log}(Z/Z_{\odot})\sim0.60$ dex  can be seen, depending on the
treatment of the convection (convective core overshooting) and the TP-AGB phase
in the stellar evolution.}}
  \label{sspmod}
\end{center}
\end{figure*}

\subsection{Improved Stellar Population \newline Models}

Today the whole extragalactic community is aware that stars in
their TP-AGB phase are responsible for more than 
half of the NIR flux that originates from a stellar population at
intermediate ages (several 100 Myr to 1-2 Gyr). In some cases, the TP-AGB
contribution may even increase to 80\%, depending on the range of
ages and metallicities adopted in the modelling (Maraston 2005).
As a consequence, AGB stars are the dominant contributors 
to the NIR mass-to-light ratio of intermediate-age stellar populations
(see Section~\ref{agb}), which are particularly sensitive for stellar mass
estimates of high-redshift galaxies at $z\sim$2 (Maraston et al. 2006).

Over the past few years, significant progress has been made in our theoretical
understanding of stellar populations (Biscardi et al. 2008; Raimondo 2009;
Conroy et al. 2009; Lee et al. 2010; Gonz{\'a}lez-L{\'o}pezlira et al. 2010;
Conroy \& Gunn 2010). Substantial improvements were obtained regarding the
uncertain and unknown late stages in the evolution of massive, evolved
stars, in particular the (post)-AGB phase and
the horizontal branch phase. A full discussion is beyond the scope of this
review. However, one of the most promising techniques using a flexible 
synthesis of stellar populations, which has been verified through detailed
comparisons with high-quality observations, is presented in the
following.

Flexible Stellar Population Synthesis (FSPS) descriptions take a novel
approach in the construction of synthetic SSP spectra (Conroy et al. 2009;
Conroy \& Gunn 2010). Through a parameterisation of uncertain stages in the
stellar evolution and by allowing these variables to alter freely, both the
theoretically motivated galaxy properties as well as different sets of
observational constraints can be modelled in combination. In particular, FSPS
allows researchers to compute SSPs for a range of
Initial Mass Functions (IMFs) and metallicities and for a variety of
assumptions regarding the morphology of the horizontal branch, the blue
straggler population, the post-AGB phase, and the location in the 
Hertzsprung-Russell diagram of the TP-AGB phase. From these SSPs,
composite stellar populations (CSPs) for a
variety of star formation histories (SFHs) and dust attenuation prescriptions
can be generated. Furthermore, the descriptions offer different choices of
isochrones and stellar libraries and provide weights to various stages of the
stellar evolution.

Conroy et al. point out that the  unmodified Padova models result in a
substantial disagreement with the observed properties of stellar clusters.  
This disagreement is amplified to a significant extent
by a large population of luminous carbon stars as predicted by the
models. Therefore, the bolometric luminosity ($L_{\rm bol}$) along the
TP-AGB phase was substantially lowered by accounting for shifts in the
effective temperature $T_{\rm eff}$ ($\Delta_T$) and $L_{\rm bol}$
($\Delta_L$) along the entire TP-AGB phase (Conroy et al. 2009). 
These applied changes are independent of metallicity and reduce the
importance and dominance (both the overall number and higher number of
more luminous O-rich stars) for all TP-AGB stars, including carbon stars.
The resulting modified Padova isochrones are in much better agreement with
both MC star clusters and post-starburst galaxies (see Figure~\ref{sspmod}).
The substantial disagreement between the \textit{unmodified} Padova
calculations and the star cluster data arises from three differences: the
usage of old inferior-quality data, the application of theoretical atmosphere
models with dust radiative transfer descriptions (Loidl et al. 2001), 
and the application of the LF of carbon stars found in the MCs. In particular,
the carbon star models are significantly bluer ($>1$~mag in $V-K$) than
observed carbon stars (R. Gautschy 2009, private communication). Further,
the new models by Conroy \& Gunn (2010) adopt observed stellar spectra
including the TP-AGB phase, which provide a better description of the
observed properties of O-rich and carbon stars, rather than synthetic
modelled spectra.

In Figure~\ref{sspmod}, a comparison between FSPS models with both
the Bag of Stellar Tracks and Isochrones (BaSTI) and modified
Padova isochrones (Conroy \& Gunn 2010) and NIR SBF measurements
within the $K$-band versus integrated $(V-I)_0$ colour
plane is given. Further, data for star clusters in the 
MCs (Gonz{\'a}lez-L{\'o}pezlira et al. 2005) and SBF data for
elliptical gala\-xies (Liu et al. 2002; Jensen et al. 2003) are included.
The left-hand panel shows the results for the modified Padova isochrones,
where\-as the right-hand panel displays the findings for the modified 
FSPS+BaSTI models. It is clearly evident that the new modified FSPS+BaSTI 
prescriptions define a much broader range in the magnitude-colour plane
than the unmodified predictions. In particular, the range of constant ages is
more pronounced in the FSPS predictions and extends to the far blue colours of
stellar MC clusters. Nevertheless, a subsample of galaxies comprises 
fainter SBF magnitudes and therefore is still poorly represented
in both the BaSTI and Padova calculations. Because of the inverse relationship
between NIR SBF luminosities and (stellar) masses, if SBF luminosities are
overestimated as in the current model predictions then the derived stellar
masses and mass-to-light ratios will be underestimated.

Many observations of elliptical galaxies provide stro\-ng evidence for an 
$\alpha$-enhancement of heavy elements in their stellar populations 
(e.g., Worthey et al. 1992; Ziegler et al. 2005). For the theoretical
prescriptions in Figure~\ref{sspmod}, solar-scaled chemical compositions were
assumed. However, recent models including an accurate description of the
(TP)-AGB-phase suggest that the level of $\alpha$-enhancement
has a minor effect on the model predictions in 
the $(V-I)$ versus F160W SBF coordinate space (Lee et al. 2010).
Nevertheless, the impact of other enrichments on the SBF luminosities
and galaxy colours has not yet been explored in detail. 

Overall, the new modifications and prescriptions such as FSPS offer a much better
representation of the $K$-band SBF luminosities and colours, thereby addressing
several limitations in the unmodified models (Conroy \& Gunn 2010): 
BaSTI isochrones predict too-blue SBF colours compared to the observations
(implying too-hot AGB and TP-AGB temperatures), whereas the Padova isochrones
provide too-bright and too-red SBF magnitudes and colours (which originate
from an overabundance of carbon stars and an overluminous
TP-AGB evolutionary phase).

So far, the current status of the theoretical modelling of SBF quantities
provides some constraints on the average metallicities of unresolved
stellar populations. For the future, a better understanding of the whole range
of stellar evolution phases is crucial to gain detailed insights into the
stellar population properties of globular clusters and galaxies.

\section{A Measure of Distance-\newline Independent Luminosities}\label{ncal}

Measuring absolute luminosities of galaxies that are distance-independent
to a high precision is important for a wide range of astronomical applications
(e.g., sizes of galaxies, black hole masses, total galaxy mass measurements
etc). Tonry and collaborators have discovered a further parameterisation of
the SBF method (T01). Moreover, this description is not only 
distance-independent but also independent from the photometric calibration
or dust extinction. The parameter $\overline{N}$ is defined as the ratio of
the total apparent flux from a galaxy to the flux provided
by the fluctuation signal. In terms of magnitudes, this (absolute luminosity)
measure is the difference between the fluctuation magnitude $\overline{m}$
and the total magnitude of the galaxy $\overline{m}_{\rm tot}$,
which corresponds to the total luminosity of the galaxy in units of the
luminosity of a typical giant star within that galaxy as:
\begin{equation}
\overline{N}=\overline{m}-\overline{m}_{\rm tot}=+2.5\,{\rm log}\left( 
\frac{L_{\rm tot}}{\overline{L}}\right).
\end{equation}
$\overline{N}$ is also referred to as the `fluctuation star count'.

\begin{figure}[!t]
\begin{center}
\includegraphics[angle=0,width=3.0in]{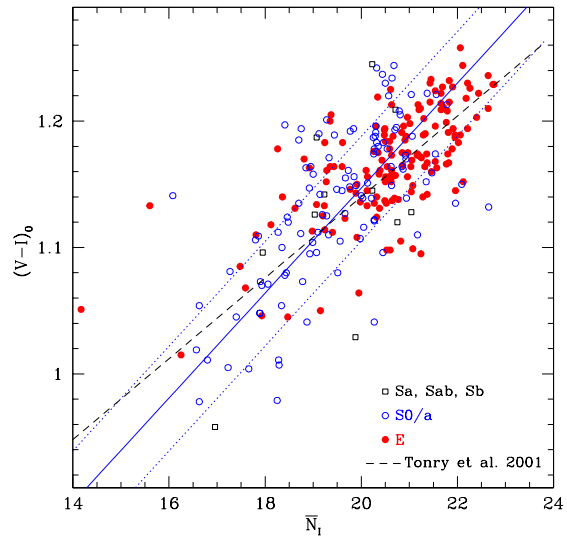}
\vspace{0.2cm}
\caption[SBF N]{\small{$(V-I)_0$  galaxy colour as a function of  
the fluctuation star count $\bar{N}_I$ for different morphological
types (Blakeslee et al. 2001b). Symbols are as in Figure~\ref{sbfhub}.
The dashed line is the empirical relation from Tonry et al. (2001).}}
\label{sbfn}
\end{center}
\end{figure}

As expected, galaxy colour correlates with $\overline{N}$ (which is a proxy for
the absolute luminosity) following the relation:
\begin{equation}
(V-I)_0= 0.317+0.042\,\overline{N}.
\label{Ncol}
\end{equation}
Figure~\ref{sbfn} shows the dependence of the $(V-I)$ galaxy colour on 
the fluctuation star count $\overline{N}$. The correlation is well established for
different morphological types (E, S0, Sa) and across a range of luminosities.
The relation in equation~\ref{Ncol}, which is shown in Figure~\ref{sbfn} as the
solid and dotted ($\pm1\sigma$ error) lines, is slightly steeper but in good
agreement with the correlation of $\overline{N}$ and $(V-I)$ found by T01
(dashed line).

Surprisingly, the slope of this relation is very shallow with an observed scatter
in $(V-I)$ of 0.04~mag. Therefore, a large error in $\overline{N}$ translates into a
negligible effect in $(V-I)$; for example assuming $\delta \overline{N}=0.5$~mag
corresponds to only $\delta(V-I)=0.016$~mag (T01). According to the prediction,
the intrinsic scatter of the relation might be as small as 0.025~mag, which also
suggests that it is an efficient way of deriving accurate extinction
measurements with rms uncertainties of 2\%.

Although there is some covariance from the application of $\overline{m_I}$
for both the distance modulus and the intrinsic galaxy colour, the actual
covariance is very mild because of the shallow slope of the $\overline{N}-(V-I)$
relationship. Instead, using the galaxy colour to estimate $\overline{M}_I$ 
directly is challenging due to the observational requirements of
high-precision photometry and the accurate assessment of the presence of
and sensitivity to dust extinction. 

%
%

Based on the $I$-band SBF survey, T01 found an empirical relationship between
$\overline{M}_I$ and $\overline{N}$ following 
\begin{equation}
\overline{M}_I=1.74+0.14\,(\overline{N}-20).
\end{equation}
Although this introduces a covariance between $\overline{M}_I$ derived in this
way and $\overline{m}_I$, the resulting distance modulus is 14\% less sensitive
to (systematic) uncertainties in $\overline{m}_I$ (Blakeslee et al. 2001b).
However, $\overline{N}$ should be not regarded as a substitute for $(V-I)$
when deriving $\overline{M}_I$, as the parameter is still relatively
unexplored. Nevertheless, in the absence of a galaxy colour, $\overline{N}$
provides an alternative way to measure a distance.
For example, Cantiello et al. (2011) have adopted a calibration based on 
$\overline{N}$ to derive the distance to 12 nearby galaxies with an accuracy
of about 30\%, where colour information was unavailable in the data archive.
The consequence of the relationship of $\overline{N}$ with $\overline{M}_I$
is suggested to rely on the Fundamental Plane of early-type galaxies  
(Djorgovski \& Davis 1987; Dressler et al. 1987) and their projections
(e.g., Fritz et al. 2005; 2009a). 
If a colour measurement is available, $\overline{N}$ should not supplant
the application of $(V-I)$ to calibrate $\overline{M}_I$. Nevertheless, it
appears that $\overline{N}$ offers a valuable alternative to measure a fairly
reliable distance, but more work is needed to understand the parameter
and its relatively large scatter fully.

\section{Conclusions}\label{con}

Surface Brightness Fluctuations are most likely the most efficient
technique to measure the distance to unresolved nearby
stellar systems: In terms of accuracy, depth, and completeness, the SBF
technique represents a unique way of measuring extragalactic distances
to early-type galaxies and spiral bulges.
Moreover, as SBFs are an intrinsic property of an integrated stellar
population, they give insight into and allow scientists to set constraints on
the stellar populations and chemistry of globular clusters and early-type
galaxies. Precise extragalactic distances are still extremely important and 
fundamental for various applications in astrophysics. High-accuracy
distance measurements are essential because errors in distance usually
translate into large uncertainties in the derived physical properties
of individual galaxies, such as absolute luminosities, linear sizes, star
formation rates and timescales, black hole masses, and (total) masses of
galaxies, as well as dark matter halo mass estimates. Moreover, for the nearby
universe, accurate distances serve as a reliable and trustworthy substitute for
redshift as a distance indicator and provide the possibility of mapping the
three-dimensional structure and velocity field of the local volume. 

Multi-wavelength SBF studies involving optical and NIR observations
offer a unique way of analysing the properties and content of composite
stellar populations, as different photometric bandpasses are sensitive to 
different phases of the stellar evolution. In particular, 
stellar population gradients based on SBF measurements allow one to probe the
merger and assembly history of nearby galaxies and unresolved stellar systems.
An interesting byproduct of detailed SBF data are possible constraints on the
stellar populations and properties of globular cluster systems, such as
understanding the nature and origin of the bi/multi-modal globular cluster
colour distributions in giant elliptical galaxies, which remains an unsolved
puzzle (Conroy \& Spergel 2011).

Apart from ground-based observations, a complete optical SBF survey from
space will allow us to construct a complete map of the individual velocity
patterns of galaxies and galaxy clusters down to an accuracy in distances
where the bulk motions become just a few percent ($\la$3\%) of the Hubble
flow (see Section~\ref{cal} for details). Based on WFC3/IR observations of
600 Cepheids to calibrate the magnitude-redshift relation of 253 SNe Ia, the
most recent measurement of the Hubble constant was derived down to an
accuracy of 3\%:
$H_{0}=73.8\pm2.4({\rm stat})\pm3.4({\rm sys})$~\kmsM (Riess et al. 2011).
Future observations will be able to measure the Hubble constant to a precision
of 2\% using the most accurate distance indicators. The extension of SBF
studies with WFC3/IR will be a highly valuable asset from this perspective,
both for the stellar content and for precise mass estimates.

The combination of a precise direct measurement of the Hubble constant
$H_{0}$ with constraints from the CMB provides a powerful test of different
models of dark energy (Hu 2005). In particular, the two independent approaches
together allow the investigation of the evolution of cosmic energy density
$w$ to 10\% accuracy across a broad cosmic epoch of $0<z<1100$ into
the early universe of the `dark ages' ($\sim$13.5 Gyr). However, $w$ is highly
susceptible to uncertainties in the value of $H_{0}$, with the errors in $w$
being twice the fractional errors in $H_{0}$, thereby keeping $w$ and the
flatness constant (Weinberg et al. 2012). The real challenge for future studies
is therefore an accurate measurement of the local $H_{0}$ value to 1\% precision,
encompassing both statistical (random) errors \textit{and} systematic
uncertainties.

The drawback of an individual $H_{0}$ value
is the restriction to a single number at a fixed redshift, so that only one
particular model of dark energy can be tested but not possible 
deviations from the theoretical predictions. A high-precision measurement of 
$H_{0}$, however, significantly increases the power of other dark energy
constraints by up to 40\% (Weinberg et al. 2012). Moreover, a direct $H_{0}$
determination has the potential to uncover possible departures from the smooth
evolution of dark energy at low redshift and could settle the controversy if
the universe is \textit{still} in a phase of acceleration.

\section{Future Directions}\label{fut}

There is no doubt that the future of SBF lies in a combination of
space-based and Adaptive Optics (AO) observations. The SBF fluctuation signal
in the $H$-band is a factor of 30 larger than that in the $I$-band. Equally
important, for {\em HST}/NICMOS observations the sky background is more than
a factor of 100 lower compared to IR observations with ground-based
telescopes. The {\em HST} WFC3/IR channel offers a unique tool with high
spatial resolution (FWHM=$0.13''$),
wide field of view ($2.05'\times2.26'$), good set of filters for SBF
measurements, and good detector characteristics (e.g., sensitivity, noise,
cosmetics, etc). WFC3/IR observations in F110W (1.25~$\mu$m, $\approx$$J$) and
F160W (1.6~$\mu$m, $\approx$$H$) as well as ACS $z'$-band (0.85~$\mu$m)
measurements will be useful complements to existing (ground-based) studies. 

The most promising approach is a complete space-based SBF IR
survey to map out the nearby universe. Such a project would allow scientists
to constrain the stellar population and mass content of `dynamically hot'
galaxies and act as a powerful complement to ongoing surveys studying 
the properties of nearby galaxies in detail, 
such as SAURON (de Zeeuw et al. 2002) or ATLAS$^{3D}$ (Cappellari et al.
2011)\footnote[8]{ATLAS$^{3D}$:~http://www-astro.physics.ox.ac.uk/atlas3d}.
A crucial complement to space observations is offered by AO observations
with NIR instruments [e.g., ALTtitude conjugate Adaptive optics for the
InfraRed (ALTAIR), OH-Suppressing InfraRed Integral-field Spectrograph
(OSIRIS), Nasmyth Adaptive Optics System Near-Infra\-red Imager and Spectrograph
(NACO), InfraRed Camera and Spectrograph (IRCS)] with excellent seeing
conditions between $\la0\farcs03-0\farcs05$ FWHM in the IR wavelength
regime, feasible with current state-of-the-art 8-m class telescopes such as
Gemini, Keck, VLT, or Subaru. Future 30-m class telescopes such as the
E-ELT\footnote[9]{E-ELT:~http://www.eso.org/public/teles-instr/e-elt}
or TMT\footnote[10]{TMT:~http://www.tmt.org} 
will be operated consecutively using AO instruments.

Multi-band SBF studies across a broad wavelength regime will be the ultimate
key to understand the stellar populations of stellar systems and their
properties. If it is possible to conceive completely the dependence of
$\overline{M}$ across the full wavelength spectrum, the SBF method will then
be independent from the calibration of the primary distance indicators of
Cepheids.

The future {\em JWST} will open a new window for SBFs in the MIR, which is 
currently prohibitively expensive and challenging from the perspective of
ground-based observations (telescope exposure time and uncertainties in the
bright sky background). Promising wavelength ranges are the $L$ (3.5~$\mu$m)
and $M$ (4.8~$\mu$m) bands, as both are expected to be highly sensitive to
the age content of old, evolved stellar populations. In particular, the
MIR SBFs will allow astronomers to constrain the mass-loss rates of AGB stars
and pin down whether there is a relation between the mass-loss and metallicity
of a stellar population as expected from stellar synthesis modelling
(Gonz{\'a}lez-L{\'o}pezlira et al. 2010).

Moreover, the {\em Gaia} astrometry
satellite,\footnote[11]{Gaia satellite:~http://gaia.esa.int/} expected to launch in
May 2013, will have a dramatic impact through a re-definition of the cosmic
distance scale. Proper motions for stars and new trigonometric parallaxes of
ten-thousand (partly long-period) Cepheids will be measured within 5 kpc; hence
geometric distances of the most powerful `primary' distance indicators
(e.g., Cepheids, RR Lyrae stars, statistical/tri\-go\-no\-me\-tric parallaxes)
will be improved with unprecedented accuracy (between 10 to 100~$\mu$arcsec,
$\sim 25\mu$arc\-sec at $V=15$~mag). Distances with precision at 
sub-percentage-level will pave the way for a direct
refinement of the calibration of `secondary' distance indicators (e.g., SBF,
SNe Ia/II, Novae, $D_n-\sigma$, Tully-Fisher relation) and a global
re-assessment of the entire cosmic distance ladder (see also 
Section~\ref{cont}). Moreover, due to the frequent repeated observations
of the same areas on the sky, {\em Gaia} will also detect and locate transient
events, such as gamma-ray-bursts (GRBs), binaries and SNe with a completeness
down to $G=20$~mag, to a high accuracy. 

The combination of both high-resolution trigonometric parallaxes and
NIR-AO observations will lift the SBF technique to the next level,
establish an ultimate reference point for the cosmic distance scale,
and settle the controversy regarding the local Hubble flow
(Lynden-Bell et al. 1988; Tonry et al. 2000).
The future of the SBF method still shines extremely brightly.
 
\section*{Acknowledgments} 

The author is grateful to Professor Joseph~B. Jensen for stimulating discussions and
many valuable comments and suggestions on an earlier version of the draft. The
anonymous referee is thanked for a constructive review, which improved the
clarity of the manuscript. The author acknowledges support from a VIPERS 
Fellowship through a PRIN-INAF 2008 grant (VIPERS) and partial support from
grant HST-GO-10826.01 from the Space Telescope Science Institute, which is
operated by the Association of Universities for Research in Astronomy,
Inc., under NASA contract NAS 5-26555.


\end{document}